\documentclass[skeleton]{SciPost}

\usepackage{amsmath,amssymb}
\usepackage{braket}
\usepackage{bm}
\usepackage[mathscr]{euscript}
\usepackage{enumitem}
\usepackage{amsfonts}
\usepackage{graphicx}
\usepackage{hyperref}

\usepackage{hyperref}
\hypersetup{
  colorlinks = true,
  citecolor = blue,
  urlcolor = blue
}

\newcommand{{\bfl}}{\mbox{\boldmath$l$\unboldmath}}
\newcommand{{\bfx}}{\mbox{\boldmath$x$\unboldmath}}
\newcommand{{\bfk}}{\mbox{\boldmath$k$\unboldmath}}
\newcommand{{\bfv}}{\mbox{\boldmath$v$\unboldmath}}
\newcommand{{\bfq}}{\mbox{\boldmath$q$\unboldmath}}
\newcommand{{\bfp}}{\mbox{\boldmath$p$\unboldmath}}
\newcommand{{\bfa}}{\mbox{\boldmath$a$\unboldmath}}
\newcommand{{\bfT}}{\mbox{\boldmath$T$\unboldmath}}
\newcommand{{\bfE}}{\mbox{\boldmath$E$\unboldmath}}
\newcommand{{\bfm}}{\mbox{\boldmath$m$\unboldmath}}
\newcommand{{\bfe}}{\mbox{\boldmath$e$\unboldmath}}
\newcommand{{\bfc}}{\mbox{\boldmath$c$\unboldmath}}
\newcommand{{\bfg}}{\mbox{\boldmath$g$\unboldmath}}
\newcommand{{\bff}}{\mbox{\boldmath$f$\unboldmath}}
\newcommand{{\bfF}}{\mbox{\boldmath$F$\unboldmath}}
\newcommand{{\bfA}}{\mbox{\boldmath$A$\unboldmath}}
\newcommand{{\bfB}}{\mbox{\boldmath$B$\unboldmath}}
\newcommand{{\bfS}}{\mbox{\boldmath$S$\unboldmath}}
\newcommand{{\bfL}}{\mbox{\boldmath$L$\unboldmath}}
\newcommand{{\bfR}}{\mbox{\boldmath$R$\unboldmath}}
\newcommand{{\bfC}}{\mbox{\boldmath$C$\unboldmath}}
\newcommand{{\bfcalO}}{\mbox{\boldmath$\cal O$\unboldmath}}
\newcommand{{\bfsigma}}{\mbox{\boldmath$\sigma$\unboldmath}}
\newcommand{{\bfmu}}{\mbox{\boldmath$\mu$\unboldmath}}

\newcommand{{\bftheta}}{\mbox{\boldmath$\theta$\unboldmath}}
\newcommand{{\bfdelta}}{\mbox{\boldmath$\delta$\unboldmath}}
\newcommand{{\bfphi}}{\mbox{\boldmath$\phi$\unboldmath}}
\newcommand{{\bfrho}}{\mbox{\boldmath$\rho$\unboldmath}}
\newcommand{{\bfcalB}}{\mbox{\boldmath$\cal B$\unboldmath}}
\newcommand{{\bfcalE}}{\mbox{\boldmath$\cal E$\unboldmath}}
\newcommand{{\bfcalJ}}{\mbox{\boldmath$\cal J$\unboldmath}}

\newcommand{{\bfcalR}}{\mbox{\boldmath$\cal R$\unboldmath}}
\newcommand{{\bfbeta}}{\mbox{\boldmath$\beta$\unboldmath}}

\newcommand{\barlambda}{{\mkern0.75mu\mathchar '26\mkern -9.75mu\lambda}}

\def\v#1{{\bf#1}}

\begin{document}

\begin{center}{\Large \textbf{The quantum phase of a dyon\\}}\end{center}

\begin{center}
	Ricardo Heras\textsuperscript{$\star$}
\end{center}

\begin{center}
	School of Physical Sciences, The Open University, Walton Hall, Milton Keynes MK7 6AA, UK
	\\
	${}^\star${\small \sf ricardo.heras@ou.ac.uk}
\end{center}

\begin{abstract}
\noindent When a dyon encircles an infinitely-long solenoid enclosing uniform electric and magnetic fields, its wave function accumulates a duality-invariant quantum phase, which is topological because it depends on a winding number and is nonlocal because the enclosed fields act on the dyon in regions where these fields vanish. Here, we derive this dyon phase and show how its duality symmetry unifies the Aharonov-Bohm phase with its dual phase. We obtain the energy levels, the two-slit interference shift, and the scattering amplitude associated with the duality-invariant quantum phase. Assuming that the dyon has spin 1/2, we show that this spin does not affect the introduced phase. We argue that a spin 1/2 dyon has electric and magnetic moments, the former being greater than the latter because of the Schwinger-Zwanziger quantisation condition.
\end{abstract}


\section*{\large 1 Introduction}
In quantum mechanics it is well-known that the wave function of a particle having the electric charge $q$ and encircling an infinitely-long solenoid enclosing a uniform magnetic flux $\Phi_m$ accumulates the Aharonov-Bohm (AB) phase \cite{1}: $\delta_{\rm AB}\!=\!nq\Phi_m/(\hbar c)$, which is topological because it depends on the winding number $n$ and is nonlocal because the magnetic field of the solenoid acts on the electric charge in regions where this field is excluded. Less known is the fact that quantum mechanics also predicts that the wave function of a particle having a magnetic charge $g$ (a magnetic monopole) and encircling an infinitely-long solenoid enclosing a uniform electric flux $\Phi_e$ accumulates the dual of the Aharonov-Bohm (DAB)\footnote{Several physicists often refer to the Aharonov-Casher phase \cite{4}: $\delta_{\rm AC}=4\pi \mu\lambda_{e}/(\hbar c),$ in which a magnetic moment of magnitude $\mu$ moves around, and parallel to, an infinitely long rod possessing the linear electric charge density $\lambda_{e},$ as the dual of the AB phase (see, for example, Ref.~\cite{5}). However, this affirmation is formally incorrect because there is no consistent electromagnetic duality between electric charges and magnetic dipoles.} phase \cite{2,3}: $\delta_{\rm DAB}\!=\!-ng\Phi_e/(\hbar c)$, which is also topological because it depends on the winding number $n$ and is nonlocal because the electric field of the solenoid acts on the magnetic charge in regions where this field vanishes. If we make the dual changes: $q\!\to\! g$ and $\Phi_m\!\to -\Phi_e$ into the AB phase then we obtain the DAB phase.

Given the phases $\delta_{\rm AB}$ and $\delta_{\rm DAB}$, we could envision a more general phase $\delta_{\rm D}$ by simply adding both phases: $\delta_{\rm D}=\delta_{\rm AB}+ \delta_{\rm DAB}$, or in more explicit terms,
\begin{equation}
\delta_{\rm D}=\frac{n}{\hbar c}(q\Phi_m-g\Phi_e).
\end{equation}
The picture of the electromagnetic configuration associated with the envisioned phase $\delta_{\rm D}$ would consist of a dyon \cite{6,7,8} with electric charge $q$ and magnetic charge $g$ encircling an infinitely-long solenoid enclosing a uniform magnetic flux $\Phi_m$ and a uniform electric flux $\Phi_e$ (see Fig.~1).
\begin{figure}
  \centering
  \includegraphics[width= 137pt]{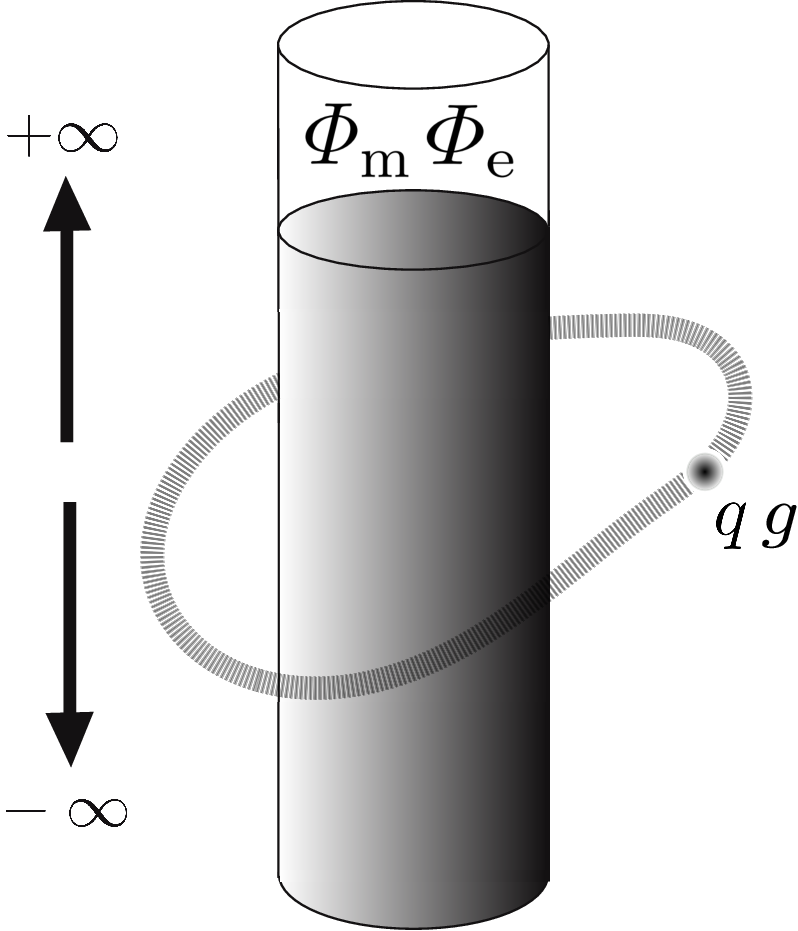}
  \caption{\small A dyon encircling a dual solenoid enclosing electric and magnetic fluxes accumulates a duality-invariant quantum phase which is topological and nonlocal.}\label{Fig1}
\end{figure}

Put in other words, one could conjecture that quantum mechanics also predicts the existence of the phase $ \delta_{\rm D}$, which has the theoretical appeal of being duality invariant, i.e., the phase $\delta_{\rm D}$ is invariant under the dual changes:  $q\to g, g\to -q, \Phi_m\to -\Phi_e$ and
$\Phi_e\to \Phi_m$, or more generally, under the $U(1)$ duality transformations: $q+ig={\rm e}^{-i\theta}\big(q'+ig'\big)$ and $\Phi_{e}+i\Phi_{m}={\rm e}^{-i\theta}\big(\Phi'_{e}+i\Phi'_{m}\big),$ where $\theta$ is an arbitrary angle.

On the other hand, the existence of the phase $\delta_{\rm D}$ is supported by semi-classical considerations based on the recently discussed correspondence between topological electromagnetic quantum phases and topological electromagnetic angular momenta \cite{9,10}. It has been shown \cite{9} that the configuration formed by an electric charge $q$ encircling an infinitely-long solenoid enclosing a uniform magnetic flux $\Phi_m$ accumulates an electromagnetic angular momentum  $L_q=n q \Phi_m/(2\pi c)$, which can be considered as the classical counterpart of the AB phase: $\delta_{\rm AB}=nq\Phi_m/(\hbar c)$, being both quantities connected by\footnote{To our knowledge, the relation $\delta_{\rm AB}= 2\pi L_q/\hbar$ was first pointed out by Maeda and Shizuya \cite{11} and subsequently by Wakamatsu et al. \cite{12}.} $\delta_{\rm AB}= 2 \pi L_q/ \hbar.$ Similarly, it has  been shown \cite{10} that the configuration formed by a magnetic charge $g$ encircling an infinitely-long solenoid enclosing a uniform electric flux $\Phi_e$ accumulates an electromagnetic angular momentum $L_g=-n g \Phi_e/(2\pi c)$ and this can be considered as the classical counterpart of the DAB phase $\delta_{\rm DAB}=-ng\Phi_e/(\hbar c)$, being both quantities connected by $\delta_{\rm DAB}=2\pi L_g/ \hbar.$ Moreover, it was also demonstrated \cite{10} that a dyon encircling an infinitely-long solenoid enclosing uniform electric and magnetic fields accumulates the electromagnetic angular momentum $L_{qg}=n(q\Phi_m-g\Phi_e)/(2\pi c)$. Considering the relations $\delta_{\rm AB}= 2 \pi L_q/ \hbar$ and $\delta_{\rm DAB}=2\pi L_g/ \hbar$ we can conjecture the validity of the relation $\delta_{\rm D}= 2\pi L_{qg}/\hbar,$ which would imply the existence of the phase $\delta_{\rm D}$ given in Eq.~(1).

However, the formal quantum mechanical derivation of the duality-invariant phase $\delta_{\rm D}$ is not as simple as it might seem at first sight for the following reason: while the Hamiltonians associated to the Schr\"odinger equations that predict the phases $\delta_{\rm AB}$ and $\delta_{\rm DAB}$ are easy to construct, the Hamiltonian corresponding to the Schr\"odinger equation that would predict the duality-invariant phase $\delta_{\rm D}$ is not easy to elucidate because there is no canonical (local) Lagrangian formulation of a dyon interacting with given electric and magnetic fields \cite{13,14}.
But this well-known problem does not preclude us to consider the idea of a nonlocal Lagrangian formulation with the purpose to derive the phase $\delta_{\rm D}$.

In this paper we suggest a nonlocal Lagrangian for a dyon interacting with the electric and magnetic fields of an infinitely-long ``dual'' solenoid enclosing a uniform magnetic flux $\Phi_m$ and a uniform electric flux $\Phi_e$. We make use of the associated Hamiltonian to construct the corresponding Schr\"odinger equation associated with a dyon in the region outside the dual solenoid. We then solve this equation and show how the wave function of a dyon encircling the dual solenoid accumulates the duality-invariant quantum phase $\delta_{\rm D}$. We point out that this phase is topological because it depends on the winding number $n$ and is nonlocal because the fields of the dual solenoid act on the dyon in a region where these fields are zero. We explicitly demonstrate how the duality symmetry of the phase $\delta_{\rm D}$ provides a unified model of the phases $\delta_{\rm AB}$ and $\delta_{\rm DAB}$ and suggest different physical interpretations of the AB phase. We then obtain the energy levels, the two-slit interference shift, and the scattering amplitude associated with the phase $\delta_{\rm D}$. Next, we discuss the role of the spin of the dyon and suggest that if the dyon is a fermion of spin 1/2 then the phase $\delta_{\rm D}$ is insensitive to the presence of the dyon spin. Finally, we argue that a spin 1/2 dyon has electric and magnetic moments and that former is greater than the latter when considering the Schwinger-Zwanziger quantisation condition.

\section*{\large 2 The potentials of a dual solenoid}
Consider an infinitely-long dual solenoid of radius $R$ that confines its electric and magnetic fields. We choose the $z$-axis to be the axis of the dual solenoid and adopt Gaussian units as well as cylindrical coordinates $(\rho,\phi,z)$. The classical electromagnetism of this configuration is discussed in detail in Ref.~\cite{10}. We will present here only the main results. The magnetic and electric fields are given by
\begin{equation}
\v B = \frac{\Phi_{\rm m} \Theta(R-\rho)}{\pi R^2}\hat{\v z},\quad \v E = \frac{\Phi_{\rm e} \Theta(R-\rho)}{\pi R^2}\hat{\v z},
\end{equation}
where $\Phi_{\rm m}=\pi R^2 B$ and $\Phi_{\rm e}=\pi R^2 E$ are the magnetic and electric fluxes through the dual solenoid, with $B$ and $E$  being the magnitudes of the fields inside the dual solenoid, and $\Theta$ is the Heaviside step function. Inside the dual solenoid ($\rho<R$) we have $\Theta=1$ and the fields take the constant values $\v B_{\rm in}= \Phi_{\rm m}\hat{\v z}/(\pi R^2)$ and $\v E_{\rm in}= \Phi_{\rm e}\hat{\v z}/(\pi R^2)$ while outside the dual solenoid ($\rho>R$) we have $\Theta=0$ and the fields vanish $\v B_{\rm out}=0$ and $\v E_{\rm out}=0$. At the surface of the dual solenoid ($\rho=R$) the fields are not defined  due to the step function discontinuity. However, an appropriate regularisation yields $\v B(R)=0$ and $\v E(R)=0$, indicating that these fields are continuous at $\rho=R$ \cite{10}. The fields of the dual solenoid are connected with their corresponding potentials by the relations  $\v B = \nabla \times \v A$ and $\v E = -\nabla\times \v C,$ where $\v A$ is the magnetic vector potential and $\v C$ is the electric vector potential. If we adopt the Coulomb gauge conditions $\nabla \cdot \v A =0$ and $\nabla \cdot \v C=0$ these potentials can be written as \cite{10}
\begin{equation}
\v A = \frac{\Phi_{\rm m}}{2\pi}\bigg[ \frac{\Theta(\rho-R)}{\rho} + \frac{\rho \Theta(R-\rho)}{R^2} \bigg]\,\hat{\!\bfphi}, \quad \v C = -\frac{\Phi_{\rm e}}{2\pi}\bigg[ \frac{\Theta(\rho-R)}{\rho} + \frac{\rho \Theta(R-\rho)}{R^2} \bigg]\,\hat{\!\bfphi}.
\end{equation}
Inside the dual solenoid ($\rho<R$) we have $\Theta(\rho-R)=0$ and $\Theta(R-\rho)=1$ and the potentials in Eq.~(3) reduce to $\v A_{\rm in}=\rho\Phi_{\rm m}/(2\pi R^2)\,\hat{\!\bfphi}$ and  $\v C_{\rm in}=-\rho\Phi_{\rm e}/(2\pi R^2)\,\hat{\!\bfphi}$. These potentials satisfy the relations $\nabla\times \v A_{\rm in}=\Phi_{\rm m}\hat{\v z}/(\pi R^2)$ and $-\nabla\times \v C_{\rm in}=\Phi_{\rm e}\hat{\v z}/(\pi R^2)$ in agreement with the expressions for the constant magnetic and electric fields inside the dual solenoid. Outside the dual solenoid ($\rho>R$) we have $\Theta(\rho-R)=1$ and $\Theta(R-\rho)=0$ and the potentials in Eq.~(3) reduce to
\begin{equation}
\v A_{\rm out}=\frac{\Phi_{\rm m}}{2\pi} \frac{\,\hat{\!\bfphi}}{\rho},\quad \v C_{\rm out}=-\frac{\Phi_{\rm e}}{2\pi} \frac{\,\hat{\!\bfphi}}{\rho},
\end{equation}
which satisfy  $\nabla\times \v A_{\rm out}=0$ and $-\nabla\times \v C_{\rm out}=0$ in agreement with the value of the fields outside the dual solenoid. At the surface of this solenoid ($\rho=R$), the potentials are not defined due to the discontinuity of the step function. However, after an appropriate regularisation it can be shown that  $\v A(R)=\Phi_{\rm m}\hat{\bfphi}/(2\pi R)$ and  $\v C(R)=-\Phi_{\rm e}\hat{\bfphi}/(2\pi R),$ indicating that the potentials are continuous at $\rho=R$ \cite{10}. Since the potentials $\v A_{\rm out}$ and  $\v C_{\rm out}$ are irrotational, then they can be expressed as
\begin{equation}
\v A_{\rm out}=\nabla\chi,\quad \v C_{\rm out} = \nabla\xi,
\end{equation}
where $\xi=-\Phi_{\rm e}\phi/(2\pi)$ and $\chi=\Phi_{\rm m}\phi/(2\pi)$ are multi-valued functions, i.e., the functions $\xi$ and $\chi$ violate the Schwarz integrability condition according to which the crossed second partial derivatives applied to $\xi$ and $\chi$ do not commute: $(\partial^{i}\partial^{j}-\partial^{j}\partial^{i})\xi \neq 0$ and  $(\partial^{i}\partial^{j}-\partial^{j}\partial^{i})\chi \neq 0,$ where index notation has been adopted and summation of repeated indices is understood (see, for example, Refs.~\cite{15} and \cite{16} for a discussion of this criteria of multi-valued functions in the context of the AB effect and the Dirac monopole, respectively).

\section*{\large 3 Hamiltonian formulation}
The classical interaction of the fields of the dual solenoid with a non-relativistic dyon of mass $m$ and having the charges $q$ and $g$ is described by the generalised Lorentz force
\begin{equation}
\v F= q\bigg(\v E + \frac{\dot{\v x}}{c}\times \v B \bigg) + g\bigg(\v B - \frac{\dot{\v x}}{c}\times \v E \bigg),
\end{equation}
where $\dot{\v x}=d \v x/dt$ is the velocity of the dyon and $\v F= m \ddot{\v x}$ with $\ddot{\v x}=d^2  \v x/dt^2$ being the dyon's acceleration. When one tries to construct a Hamiltonian associated with the generalised Lorentz force, we have to face the fact that there is no canonical (local) Hamiltonian that yields the generalised Lorentz force in Eq.~(6) \cite{13,14}. This seems to be an impossibility of the standard Lagrangian and Hamiltonian treatments, which shows the limitation of these methods for certain types of forces. In other words, not all forces of the form $\v F= m \ddot{\v x}$ can be obtained from a local Lagrangian or Hamiltonian.

However, this inconvenience does not prevent us from looking for a specific nonlocal Hamiltonian that could lead to the force in Eq.~(6). With this purpose, consider the nonlocal Lagrangian
\begin{equation}
L(\v x;\v x_0, \dot{\v x})=\frac{m \dot{\v x}^2}{2} + \frac{\dot{\v x}}{c}\cdot[q\v A(\v x) + g\v C(\v x)] + \int^{\v x}_{\v x_0}[g \v B(\v x) + q \v E(\v x)]\cdot d \v x',
\end{equation}
where the fields $\v E$ and $\v B$ are the fields of the dual solenoid given by Eq.~(2) and the corresponding potentials $\v A$ and $\v C$ are given by Eq.~(3). The line integral in the Lagrangian is taken along a fixed reference point $\v x_0$ to the variable point $\v x$ representing the position of the dyon. This line integral accounts for a nonlocal term in the Lagrangian in that this term depends on the values at $\v x$ and $\v x_0.$ In Appendix A, we show that the corresponding Euler-Lagrange equations yield the force
\begin{equation}
\v F = q\bigg(\v E +  \frac{\dot{\v x}}{c}\times \v B \bigg) + g\bigg(\v B - \frac{\dot{\v x}}{c}\times \v E \bigg) + \v F_{\rm s},
\end{equation}
where $\v F_{\rm s}$ is a singular nonlocal term defined only on the surface of the dual solenoid and given by
\begin{equation}
\v F_{\rm s}= \kappa (z_0-z) \delta(\rho-R)\,\hat{\!\bfrho},
\end{equation}
where $\kappa=(q\Phi_{\rm e} + g \Phi_{\rm m})/(\pi R^2)$. This singular surface term is shown to vanish after an appropriate regularisation. First, we observe that if the dyon is not on the surface of the dual solenoid ($\rho \neq R$) then $\v F_{\rm s}=0$. If the dyon approaches infinitely close to the dual solenoid ($\rho \to R$) then $\v F_{\rm s}=0$ because $\lim \delta(R-\rho)=0$ as $\rho \to R.$ On the other hand, if the dyon is on the surface of the dual solenoid ($\rho =R$) then $\v F_{\rm s}$ becomes singular. However, we may treat this singularity with a regularisation of Eq.~(9). This is done by letting $\rho\to \rho + \varepsilon$ where $\varepsilon>0$ is an infinitesimal quantity, obtaining
\begin{equation}
\v F_{\rm s}= \kappa\,(z_0-z)\,\lim_{\varepsilon \to 0} \delta[(\rho+\varepsilon)-R]\,\hat{\!\bfrho}.
\end{equation}
When $\rho=R$ we obtain
\begin{equation}
\v F_{\rm s}= \kappa\,(z_0-z)\,\lim_{\varepsilon \to 0} \delta(\varepsilon)\,\hat{\!\bfrho}=0,
\end{equation}
because $\lim \delta(\varepsilon) =0$ as $\varepsilon \to 0.$  We conclude that singular term $\v F_{\rm s}$ vanishes in all space and therefore the nonlocal Lagrangian in Eq.~(7) yields the correct expression for the generalised Lorentz force.
 Using Eq.~(7) we obtain the canonical momentum $\partial L/\partial\dot{\v x}=\v p = m \dot{\v x} + (q\v A+g\v C)/c$ which gives $\dot{\v x}= [\v p- (q\v A+g\v C)/c]/m$. This result and $H=\v p\cdot \dot{\v x}-L$ gives the corresponding nonlocal Hamiltonian
\begin{equation}
H(\v x; \v x_0, \v p)= \frac{1}{2m}\bigg(\v p- \frac{1}{c}[q\v A(\v x) + g \v C(\v x)] \bigg)^2-\int^{\v x}_{\v x_0}\,[g\v B(\v x) + q\v E(\v x)]\cdot d \v x'.
\end{equation}
In the region outside the dual solenoid ($\rho>R$), we have $\v B_{\rm out}=0$ and $\v E_{\rm out}=0$ while $\v A_{\rm out} \neq 0$ and $\v C_{\rm out}\neq 0,$ and therefore the nonlocal Hamiltonian in Eq.~(12) reduces to the local form
\begin{equation}
H= \frac{1}{2m}\bigg(\v p- \frac{1}{c}(q\v A_{\rm out} + g \v C_{\rm out}) \bigg)^2.
\end{equation}
Interestingly, the Hamiltonian in Eq.~(12) is globally nonlocal but can be local if it is considered in the region outside the dual solenoid. The local Hamiltonian in Eq.~(13) is suitable to describe the quantum mechanics of a dyon outside the dual solenoid as we will see in the next section.

\section*{\large 4 The quantum phase of a dyon}
Consider a non-relativistic and spinless dyon encircling the dual solenoid (see Fig.~1). The canonical substitution $\v p \to - i\hbar \nabla$ in Eq.~(13) gives the corresponding time-dependent Schr\"odinger equation
\begin{equation}
i \hbar\frac{\partial \Psi}{\partial t}=\frac{1}{2m}\bigg(-i\hbar\nabla- \frac{1}{c}(q\v A_{\rm out} + g \v C_{\rm out}) \bigg)^2\Psi + V\Psi,
\end{equation}
where $V$ is a potential associated with a mechanical force that keeps the dyon encircling the dual solenoid. The nature of this force is not relevant for our analysis as long as it keeps the dyon encircling the dual solenoid.\footnote{We should note that many treatments on the AB phase do not include a potential $V$ associated with a force that keeps the particle with electric charge $q$ moving around the magnetic solenoid. This is because the inclusion of such a potential is not necessary for the derivation of the AB phase. Similarly, the derivation of the phase $\delta_{\rm D}$ does not require the inclusion of a potential $V$. However, we have included this potential in Eq.~(14) for completeness in the description of the system.} Since the vector potentials in Eq.~(14) can be written as the gradients of scalar functions as seen in Eq.~(5), then a solution of Eq.~(14) can be obtained by multiplying the solution $\Psi_0$ which satisfies Eq.~(14) when $\v A_{\rm out}=0$ and $\v C_{\rm out}=0$, by a suitable phase factor
\begin{equation}
\Psi(\v x,t)= {\rm e}^{i/(\hbar c)\int_{\Gamma}(q \v A_{\rm out} + g \v C_{\rm out})\cdot d \v x'}\Psi_0(\v x,t),
\end{equation}
where the line integral in the phase is taken along the dyon path $\Gamma$ from a fixed reference point $\small\bfcalO$ to the variable point $\v x$. Equation (15) assumes that the points $\small\bfcalO$ and $\v x$ never lie on the dual solenoid while the path $\Gamma$ never crosses it. As the dyon continuously encircles the dual solenoid, it follows that any dyon path $\Gamma$ can be decomposed as $\Gamma=C+\gamma$ where $C$ is any closed path that accounts for the number of times the dyon encircles the dual solenoid and $\gamma$ is any non-closed path that accounts for the open trajectory that the dyon takes before completing another turn around the dual solenoid. It is pertinent to say that a similar decomposition was discussed in the context of the AB phase \cite{15}. It then follows that the line integral in the phase in Eq.~(15) can be written as
\begin{equation}
\int_{\Gamma}(q \v A_{\rm out} + g \v C_{\rm out})\cdot d \v x'=\oint_{C}(q \v A_{\rm out} + g \v C_{\rm out})\cdot d \v x' + \int_{\gamma}(q \v A_{\rm out} + g \v C_{\rm out})\cdot d \v x'.
\end{equation}
To evaluate the circulation let us consider the following result demonstrated in Ref.~\cite{10}:
\begin{eqnarray}
\oint_C(\v A_{\rm out} + \v C_{\rm out})\cdot d \v x'=\left\{\begin{array}{@{}l@{\quad}l}
       n(\Phi_{\rm m}-\Phi_{\rm e}) & \mbox{if $C$ encloses the dual solenoid} \\[\jot]
      \,0 & \mbox{otherwise}
    \end{array}\right.
\end{eqnarray}
where $n$ is the winding number of the path $C$.
Using Eqs.~(16) and (17) we can write
\begin{equation}
\Psi(\v x,t)= [{\rm e}^{in(q\Phi_{\rm m}- g \Phi_{\rm e})/(\hbar c)}]{\rm e}^{i/(\hbar c)\int_{\gamma}(q \v A_{\rm out} + g \v C_{\rm out})\cdot d \v x'}\Psi_0(\v x,t),
\end{equation}
which states that after the dyon takes $n$ turns around the dual solenoid,
its wave function picks up the phase factor ${\rm e}^{in(q\Phi_{\rm m}-g\Phi_{\rm e})/(\hbar c)},$ and it accumulates the phase
\begin{equation}
\delta_{\rm D}=\frac{n}{\hbar c}(q\Phi_{\rm m} - g \Phi_{\rm e}).
\end{equation}
The phase $\delta_{\rm D}$ is topological because it depends on a winding number $n,$ which identifies with the number of times the dyon encircles the dual solenoid. The phase $\delta_{\rm D}$ also reflects a nonlocal interaction in that the fields of the dual solenoid act on the dyon in a region for which these fields are zero. A further feature of the phase $\delta_{\rm D}$ is its electromagnetic duality. This symmetry and some of its consequences will be discussed in the next section.

\section*{\large 5 Duality symmetry of the quantum phase $\bfdelta_{\bf{D}}$}
It is easy to see that the phase $\delta_{\rm D}$ in Eq.~(19) is invariant under the simultaneous application of the set of duality transformation of charges $\{q\to g,\,g\to -q\}$ and the set of duality transformations of fluxes $\{\Phi_{\rm m}\to -\Phi_{\rm e},\,\Phi_{\rm e}\to \Phi_{\rm m}\}$. We note that both sets of transformations are independent because the charges $q$ and $g$ of the dyon are specified independently from the fluxes $\Phi_{\rm e}$ and $\Phi_{\rm m}$ of the dual solenoid. Accordingly, the duality symmetry of the phase $\delta_{\rm D}$ corresponds to a dyon/dual-solenoid composite. We can generalise the discrete duality symmetry of the phase $\delta_{\rm D}$ through the set of $U(1)$ electromagnetic duality transformations given by
\begin{equation}
q+ig={\rm e}^{-i\theta}\big(q'+ig'\big),\quad\Phi_{\rm e}+i\Phi_{\rm m}={\rm e}^{-i\theta}\big(\Phi'_{\rm e}+i\Phi'_{\rm m}\big),
\end{equation}
where $\theta$ is an arbitrary real angle. The transformations in Eq.~(20) can explicitly be written as
\begin{eqnarray}
q&=&\,q'\cos\theta+g'\sin\theta,\;\quad\; \,\, \,\,\quad g=-q'\sin\theta+g'\cos\theta,\\
\Phi_{\rm e}&=&\,\Phi_{\rm e}'\cos\theta+\Phi_{\rm m}'\sin\theta,\;\;\quad \Phi_{\rm m}=-\Phi_{\rm e}'\sin\theta+\Phi_{\rm m}'\cos\theta,
\end{eqnarray}
and their corresponding inverse transformations read
\begin{eqnarray}
q'&=&\,q\cos\theta-g\sin\theta,\;\quad\;\;\,\,\,\, \,\,\quad g'=q\sin\theta+g\cos\theta,\\
\Phi'_{\rm e}&=&\,\Phi_{\rm e}\cos\theta-\Phi_{m}\sin\theta,\; \,\, \,\quad\Phi'_{\rm m}=\Phi_{\rm e}\sin\theta+\Phi_{\rm m}\cos\theta.
\end{eqnarray}
Using Eqs.~(21) and (22) we obtain the duality invariant relation
\begin{equation}
q\Phi_{\rm m}-g\Phi_{\rm e}=q'\Phi_{\rm m}'-g'\Phi_{\rm e}'.
\end{equation}
Considering this relation we can directly see that the phase $\delta_{\rm D}$ is duality invariant:
\begin{equation}
\delta_{\rm  D}= \frac{n}{\hbar c}(q\Phi_{\rm m}-g\Phi_{\rm e})= \frac{n}{\hbar c}(q'\Phi_{\rm m}'-g'\Phi_{\rm e}')=\delta'_{\rm D},
\end{equation}
which shows that the phase $\delta_{\rm D}$ is invariant under the continuous duality symmetry specified by the transformations in Eq.~(20). By exploiting the arbitrariness of the angle $\theta$, we can show that the phase $\delta_{\rm D}=n(q\Phi_{\rm m}-g\Phi_{\rm e})/(\hbar c)$ unifies the AB phase $\delta_{\rm AB}=nq\Phi_{\rm m}/(\hbar c)$ and the DAB phase $\delta_{\rm DAB}=-ng\Phi_{\rm e}/(\hbar c)$.

We have two procedures to obtain $\delta_{\rm AB}$ from $\delta_{\rm D}$. In the first procedure, we assume that all dyons have the same ratio of magnetic to electric charge: $g'/q'$= constant. Since $\theta$ is arbitrary, we can fix this angle to satisfy
\begin{equation}
\frac{g'}{q'}=\tan\theta,
\end{equation}
which implies $\theta=\tan^{-1}(g'/q').$ The condition in Eq.~(27) and the second transformation in Eq.~(21) imply the vanishing of the magnetic charge $g$ of the dyon
\begin{eqnarray}
\nonumber g= -q'\sin\theta+g'\cos\theta = -q' \sin \bigg[ \tan^{-1} \bigg(\frac{g'}{q'}\bigg) \bigg] +g'\cos\bigg[\tan^{-1}\bigg(\frac{g'}{q'}\bigg)  \bigg]\\
= -\frac{g'}{\sqrt{(g'/q')^2+1}} + \frac{g'}{\sqrt{(g'/q')^2+1}}=0, \qquad\qquad \qquad \qquad \qquad \quad\,\,\,\,\,
\end{eqnarray}
where we have used the trigonometric relation $\sin[\tan^{-1}(a/b)]=a/(b\sqrt{(a/b)^2 +1})$ together with $\cos[\tan^{-1}(a/b)]=1/\sqrt{(a/b)^2 +1}$. Using Eq.~(28) the transformations in Eq.~(23) become
\begin{equation}
q'=q\cos\theta, \quad g'=q\sin\theta.
\end{equation}
Using Eq.~(29) and Eq.~(22) it follows
\begin{equation}
q'\Phi'_{\rm m}-g'\Phi'_{\rm e}= q(-\Phi_{\rm e}'\sin\theta+\Phi_{\rm m}'\cos\theta)=q \Phi_{\rm m}.
\end{equation}
Using Eq.~(30) the phase $\delta_{\rm D}$ reduces to the phase $\delta_{\rm AB}$ for the angle specified by Eq.~(27)
\begin{equation}
\delta_{\rm D}=\bigg[\frac{n(q'\Phi'_{\rm m}-g'\Phi'_{\rm e})}{\hbar c}\bigg]_{\theta=\tan^{-1}(g'/q')}= \frac{nq \Phi_{\rm m}}{\hbar c}= \delta_{\rm AB}.
\end{equation}

\begin{figure}
  \centering
  \includegraphics[width= 123pt]{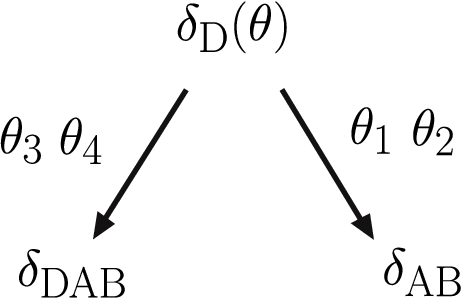}
  \caption{\small The duality symmetry of the phase $\delta_{\rm D}(\theta)$ provides a unified model of the AB and DAB phases. For the angle defined either by $\theta_1=\tan^{-1}(g'/q')$ or $\theta_2 = \tan^{-1}(-\Phi'_e/\Phi'_m)$, the phase $\delta_{\rm D}(\theta)$ becomes the AB phase $\delta_{\rm AB}$, and for the angle $\theta$ defined either by $\theta_3 = \cot^{-1}(-g'/q')$ or $\theta_4 = \cot^{-1}(\Phi'_e/\Phi'_m)$, the phase $\delta_{\rm D}(\theta)$ becomes the DAB phase $\delta_{\rm DAB}.$}\label{Fig2}
\end{figure}
In the second procedure, we assume that the ratio of the electric to magnetic fluxes is a constant quantity. Since $\theta$ is arbitrary, we can fix this angle to satisfy
\begin{equation}
\frac{\Phi'_{\rm e}}{\Phi'_{\rm m}}=-\tan\theta,
\end{equation}
which implies $\theta = \tan^{-1}(-\Phi'_{\rm e}/\Phi'_{\rm m})$. The condition in Eq.~(32) and the first transformation in Eq.~(22) imply the vanishing of the electric flux $\Phi_{\rm e}$ of the dual solenoid
\begin{eqnarray}
\nonumber \Phi_e= \Phi_e'\cos \theta+\Phi'_m\sin\theta = \Phi'_e \cos \bigg[ \cot^{-1} \bigg(-\frac{\Phi'_m}{\Phi'_e}\bigg) \bigg] +\Phi'_m\cos\bigg[\cot^{-1} \bigg(-\frac{\Phi'_m}{\Phi'_e}\bigg)  \bigg]\\
= \frac{\Phi_e'}{\sqrt{(\Phi'_e/\Phi'_m)^2+1}} -\frac{\Phi_e'}{\sqrt{(\Phi'_e/\Phi'_m)^2+1}}=0, \qquad\qquad \qquad \qquad \qquad \qquad \qquad\,\,\,\,\,\,
\end{eqnarray}
where we have used the trigonometric relation $\cos[\cot^{-1}(-a/b)]=1/\sqrt{(b/a)^2+1}$ together with $\sin[\cot^{-1}(-a/b)]=-b/(a\sqrt{(b/a)^2 +1}),$ and thus the transformations in Eq.~(24) become
\begin{equation}
\Phi'_{\rm e}=-\,\Phi_{\rm m}\sin\theta, \quad \Phi'_{\rm m}=\Phi_{\rm m}\cos\theta,
\end{equation}
which combine with the transformations in Eq.~(21) to yield the result
\begin{equation}
q'\Phi'_{\rm m}-g'\Phi'_{e}=(q'\cos\theta+g'\sin\theta)\Phi_{\rm m}=q\Phi_{\rm m}.
\end{equation}
Equation (35) implies that the phase $\delta_{\rm D}$ originates the phase $\delta_{\rm AB}$ for the angle specified by Eq.~(32)
\begin{equation}
\delta_{\rm D}=\bigg[\frac{n(q'\Phi'_{\rm m}-g'\Phi'_{\rm e})}{\hbar c}\bigg]_{\theta = \tan^{-1}(-\Phi'_{\rm e}/\Phi'_{\rm m})}= \frac{nq \Phi_{\rm m}}{\hbar c}= \delta_{\rm AB}.
\end{equation}
Following similar procedures to those used to obtain Eqs.~(31) and (36), we can show the results
\begin{eqnarray}
\delta_{\rm D}&=&\bigg[\frac{n(q'\Phi'_{\rm m}-g'\Phi'_{\rm e})}{\hbar c}\bigg]_{\theta= \cot^{-1}(-g'/q')}= -\frac{ng \Phi_{\rm e}}{\hbar e}= \delta_{\rm DAB},\\
\delta_{\rm D}&=&\bigg[\frac{n(q'\Phi'_{\rm m}-g'\Phi'_{\rm e})}{\hbar c}\bigg]_{\theta = \cot^{-1}(\Phi'_e/\Phi'_m) }= -\frac{ng \Phi_{\rm e}}{\hbar c}= \delta_{\rm DAB},
\end{eqnarray}
for the fixed angles $\theta= \cot^{-1}(-g'/q')$ and $\theta = \cot^{-1}(\Phi'_e/\Phi'_m),$ which are two angles that allow us to derive the phase $\delta_{\rm DAB}$ from the phase $\delta_{\rm D}$.

We can now draw the lessons we have learned about the electromagnetic duality of the phase $\delta_{\rm D}$. From Eqs.~(31) and (36), and (37) and (38) we can see how the $U(1)$ duality symmetry of the phase $\delta_{\rm D}$ shows its unifying property: for the angle defined by $\theta=\tan^{-1}(g'/q')$ the magnetic charge of the dyon vanishes $g=0$ and therefore the phase $\delta_{\rm D}$ reduces to the AB phase $\delta_{\rm AB}$, whereas for the angle defined by $\theta = \tan^{-1}(-\Phi'_{\rm e}/\Phi'_{\rm m})$ the electric flux of the dual solenoid vanishes $\Phi_e=0$ and the phase $\delta_{\rm D}$ also reduces to the AB phase $\delta_{\rm AB}.$ Similarly, for the angle $\theta= \cot^{-1}(-g'/q')$ the electric charge of the dyon vanishes $q=0$ and the phase $\delta_{\rm D}$ reduces to the DAB phase $\delta_{\rm DAB}$, whereas for the angle $\theta = \cot^{-1}(\Phi'_e/\Phi'_m)$ the magnetic flux of the dual solenoid vanishes and the phase $\delta_{\rm D}$ also reduces to the DAB phase $\delta_{\rm DAB}.$ Accordingly, we may write in a slightly different notation
\begin{equation}
\delta_{\rm D}(\theta) \Longrightarrow
\begin{cases}
\delta_{\rm D}(\theta_1)= \delta_{\rm AB}, & \theta_1 =   \tan^{-1}\bigg(\dfrac{g'}{q'}\bigg) \\[0.5cm]
\delta_{\rm D}(\theta_2)= \delta_{\rm AB}, & \theta_2 = \tan^{-1}\bigg(\!-\dfrac{\Phi'_e}{\Phi'_m}\bigg)\\[0.5cm]
\delta_{\rm D}(\theta_3)= \delta_{\rm DAB},& \theta_3 =\cot^{-1}\bigg(\!-\dfrac{g'}{q'}\bigg)\\[0.5cm]
\delta_{\rm D}(\theta_4)= \delta_{\rm DAB},& \theta_4 = \cot^{-1}\bigg(\dfrac{\Phi'_e}{\Phi'_m}\bigg)
\end{cases}
\end{equation}
Equation (39) shows that the phase $\delta_{\rm D}$ depends on the angle $\theta$, i.e., $\delta_{\rm D}=\delta_{\rm D}(\theta),$ but when we fix this angle for the values specified by $\theta_1$ and $\theta_2$, then this phase reduces to the AB phase $\delta_{\rm AB}$, whereas if we fix the angle for the values specified by $\theta_2$ and $\theta_3$, the phase $\delta_{\rm D}$ becomes the DAB phase $\delta_{\rm DAB}.$ This unifying property is represented in Fig.~2.

\section*{\large 6 Interpretations of the Aharonov-Bohm phase in the light of duality}
Electromagnetic duality allows different interpretations of the AB phase. To see this, consider Eq.~(29) which holds when $g=0$ and implies the beautiful relation
\begin{equation}
q=\sqrt{q'^2+g'^2}.
\end{equation}
Multiplying this equation by $n\Phi_{m}/(\hbar c)$, we obtain an alternative expression for the AB phase
\begin{equation}
\delta_{\rm AB}=n\frac{q \Phi_{\rm m}}{\hbar c}=n\frac{\sqrt{q'^2+g'^2}\, \Phi_{\rm m}}{\hbar c}.
\end{equation}
According to this expression, the AB phase can be originated either by the nonlocal action of the magnetic flux $\Phi_{\rm m}$ on the charge $q$ (first equality) or by the nonlocal action of the magnetic flux $\Phi_{\rm m}$ on a dyon having the electric charge $q'=\,q\cos\theta$ and the magnetic charge $g'=q\sin{\theta}$ (second equality). The first equality in Eq.~(41) gives the usual nonlocal interpretation of the AB phase (see, for example, Ref.~\cite{10} for a discussion of this interpretation), whereas the second equality gives a nonlocal interpretation based on electromagnetic duality.

\begin{table}[t]
\begin{center}
\def\arraystretch{2.5}
\begin{tabular}{ |c|c|c| }
\hline
Interpretation & AB phase  \\ 
\hline
Standard (nonlocal) &$\delta_{\rm AB}=n\dfrac{q\Phi_{m}}{\hbar c}$   \\[0.1 cm]
\hline
Dual 1 (nonlocal)   & $\delta_{\text{AB}}=n \dfrac{\Phi_{m}\sqrt{q'^2+g'^2}}{\hbar c}$   \\[0.1 cm]
\hline
Dual 2 (nonlocal)   & $\delta_{\text{AB}} =n \dfrac{q\sqrt{\Phi_{e}'^2+\Phi_{m}'^2}}{\hbar c}$   \\[0.1 cm]
\hline
\end{tabular}
\vskip 10pt
\caption{\small Interpretations of the AB phase supported by electromagnetic duality. In the standard nonlocal interpretation the magnetic flux $\Phi_{m}$ of the solenoid has a nonlocal action on the electric charge $q$. In a first dual interpretation the magnetic flux $\Phi_{m}$ has a nonlocal action on the electric charge $ q'\!=\,q\cos\theta$ and magnetic charge $g'\!=q\sin\theta$ of a dyon. In a second dual interpretation the fluxes $\Phi'_{m}=\Phi_{m}\cos\theta$ and $\Phi'_{e}=-\,\Phi_{m}\sin\theta$ of a dual solenoid have a nonlocal action on the electric charge $q.$ }
\end{center}
\end{table}

Analogously, from Eq.~(34), which holds when $\Phi_{\rm e}=0$, it follows the duality-invariant relation
\begin{equation}
\Phi_{\rm m}=\sqrt{\Phi_{\rm e}'^2+\Phi_{\rm m}'^2}
\end{equation}
Multiplying this equation by $nq/(\hbar c)$ it follows
\begin{equation}
\delta_{\rm AB}=n\frac{q \Phi_{\rm m}}{\hbar c}=n
\frac{q\sqrt{\Phi_{\rm e}'^2+\Phi_{\rm m}'^2}}{\hbar c}.
\end{equation}
This expression tells us that the AB phase can be originated either by the nonlocal action of the magnetic flux $\Phi_{\rm m}$ on the charge $q$ (first equality) or by the nonlocal action of the fluxes $\Phi'_{\rm e}=-\,\Phi_{\rm m}\sin\theta$ and $\Phi'_{\rm m}=\Phi_{\rm m}\cos\theta$ of a dual solenoid
on the electric charge $q$. The first equality in Eq.~(43) gives the usual nonlocal interpretation of the AB phase, whereas the second equality gives a second nonlocal interpretation supported by electromagnetic duality. The standard nonlocal interpretation of the AB phase and the interpretations supported by the second equalities in Eqs.~(41) and (43) are shown in Table 1. We note that similar interpretations hold for the phase $\delta_{\rm DAB}$.

What do we make of Eqs.~(41) and (43)? The lesson we can draw here is that electromagnetic duality shows that it is a matter of convention to speak of electric charges and magnetic fluxes, and not of dyons and dual fluxes. Conventionally, we say that a charged particle has the electric charge $q\neq 0$ and the magnetic charge $g=0,$ but with equal right, we can say that a charged particle has the electric charge $q'=q\cos\theta$ and the magnetic charge $g'=q\sin\theta$ with $\theta$ specified by Eq.~(27). Analogously, we say conventionally that a flux tube is specified by the magnetic flux $\Phi_m \neq 0$ and the electric flux $\Phi_e =0,$ but with equal right, we can say that a flux tube has the magnetic flux $\Phi'_m=\Phi_m\cos\theta$ and the electric flux $\Phi'_e=-\Phi_m\sin\theta$ with $\theta$ specified by Eq.~(32). In this regard, duality allows equivalent descriptions of the same theory.

\section*{\large 7 Energy levels, two-slit interference shift, and scattering amplitude}
The similarity between the Hamiltonian in Eq.~(13) and the AB Hamiltonian due to a particle with charge $q$ and mass $M$ propagating outside an infinitely-long solenoid can be exploited to obtain some relevant quantities associated with the phase $\delta_D.$ To see this, consider the AB Hamiltonian operator
\begin{equation}
\widehat{H}_{\rm AB}= \frac{1}{2M}\bigg(-i \hbar \nabla - \frac{q}{c}\v A_{\rm out}\bigg)^2+V.
\end{equation}
This operator can be written as
\begin{equation}
\widehat{H}_{\rm AB}= \frac{1}{2M}\bigg(-i \hbar \nabla - L_q \nabla\phi\bigg)^2+V,
\end{equation}
where $L_q=q\Phi_{\rm m}/(2\pi c)$ is the electromagnetic angular momentum of the configuration formed by a charge encircling an infinitely-long solenoid \cite{9}. Consider now the Hamiltonian operator
\begin{equation}
\widehat{H}_{\rm D}= \frac{1}{2m}\bigg(-i \hbar \nabla - \frac{1}{c}(q\v A_{\rm out} + g \v C_{\rm out})\bigg)^2+V,
\end{equation}
which corresponds to the Schr\"odinger equation in Eq.~(14). We can write this operator as
\begin{equation}
\widehat{H}_{\rm D}= \frac{1}{2m}\bigg(-i \hbar \nabla - L_{qg} \nabla\phi\bigg)^2+V,
\end{equation}
where $L_{qg}=(q\Phi_{\rm m}-g\Phi_{\rm e})/(2\pi c)$ is the electromagnetic angular momentum of the configuration formed by a dyon encircling an infinitely-long dual solenoid \cite{10}. We observe that $\widehat{H}_{\rm AB}$ in Eq.~(44) and $\widehat{H}_{\rm D}$ in Eq.~(47) have the same form modulo that the former involves $L_{q}$ and $M$ and the latter $L_{qg}$ and  $m$. It then follows that we can obtain the quantum effects corresponding to the operator $\widehat{H}_{\rm D}$ from the quantum effects associated with the operator $\widehat{H}_{\rm AB}$ by making the replacements
\begin{equation}
L_{q}\to L_{qg},\quad M\to m.
\end{equation}
These replacements provide a simple approach through which we can obtain quantities associated to the dyon phase $\delta_{\rm D}$ from quantities associated to the AB phase $\delta_{\rm AB}$. Here we will use this method in three specific examples: (i) the energy levels of a dyon encircling the dual solenoid, (ii) the two-slit interference shift due to dyons propagating outside the dual solenoid, and (iii) the scattering amplitude due to dyons propagating outside the dual solenoid. We will then see how the results (i)-(iii) hold in the limit of a vanishing radius of a dual solenoid, i.e., a dual flux line.

\begin{figure}
  \centering
  \includegraphics[width= 128pt]{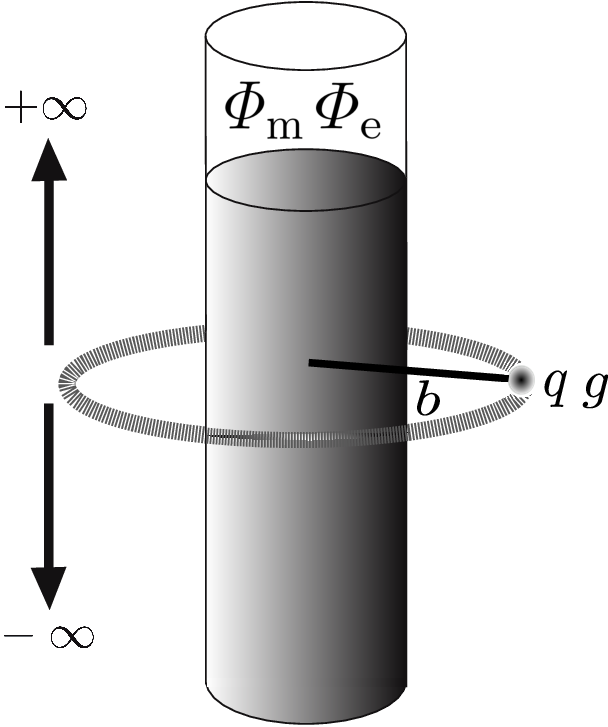}
  \caption{{\small The energy levels of the configuration formed by a dyon encircling an infinitely-long solenoid whose motion is constrained in a circle of fixed radius $b$ depends on the confined electric and magnetic fluxes even though the dyon lies in a region inaccessible to these fluxes.}}\label{Fig3}
\end{figure}
\subsection*{\normalsize 7.1 Energy Levels}
\noindent The energy levels of the configuration formed by an electric charge encircling an infinitely-long solenoid of radius $R$ and whose motion is constrained in a circle of fixed radius $b$ are well-known and can be expressed in terms of its associated electromagnetic angular momentum as
\begin{equation}
E_{\mu}= \frac{\hbar^2}{2 M b^2}\bigg(\mu - \frac{L_q}{\hbar}\bigg)^2,
\end{equation}
where $\mu$ is an integer. If we make the substitutions in Eq.~(48) into Eq.~(50) and insert explicitly $L_{qg}= (q\Phi_{m}-g\Phi_{e})/(2\pi c)$ then we obtain the energy levels
\begin{equation}
E_{\mu}= \frac{\hbar^2}{2 m b^2}\bigg(\mu - \frac{(q \Phi_{m}- g\Phi_{e})}{2\pi \hbar c}\bigg)^2,
\end{equation}
which correspond to the configuration formed by a dyon encircling a dual solenoid (see Fig.~3).

\subsection*{\normalsize 7.2 Two-slit interference shift}
\noindent Consider a two-slit interference effect in which identical particles having the electric charge $q$ propagate from a source, pass through two slits on a first screen and are detected on a second screen. If an infinitely-long solenoid that confines its magnetic field is placed between the two screens then there is an extra shift detected on the second screen. This shift has been calculated in detailed by Kobe \cite{17} and in terms of the electromagnetic angular momentum can be written as
\begin{equation}
\Delta_{\rm AB}=\frac{2\pi L \barlambda}{d \hbar}L_q,
\end{equation}
where $d$ is the distance between the two slits on the first screen, $L$ is the distance between the two screens, and $\barlambda= \hbar/(M v)$ is the de Broglie wavelength of the electrically charged particle having the velocity $v$. If we make the substitutions in Eq.~(48) into Eq.~(49) and insert explicitly $L_{qg}= (q\Phi_{m}-g\Phi_{e})/(2\pi c)$, then we obtain the shift
\begin{equation}
\Delta_{\rm D}=\frac{L \barlambda}{d}\frac{(q \Phi_m - g \Phi_e)}{\hbar c},
\end{equation}
which correspond to the two-slit interference shift due to identical dyons propagating outside a dual solenoid (see Fig.~4), where now $\barlambda= \hbar/(mv)$ is the de Broglie wavelength of the dyon.
\begin{figure}
  \centering
  \includegraphics[width= 309pt]{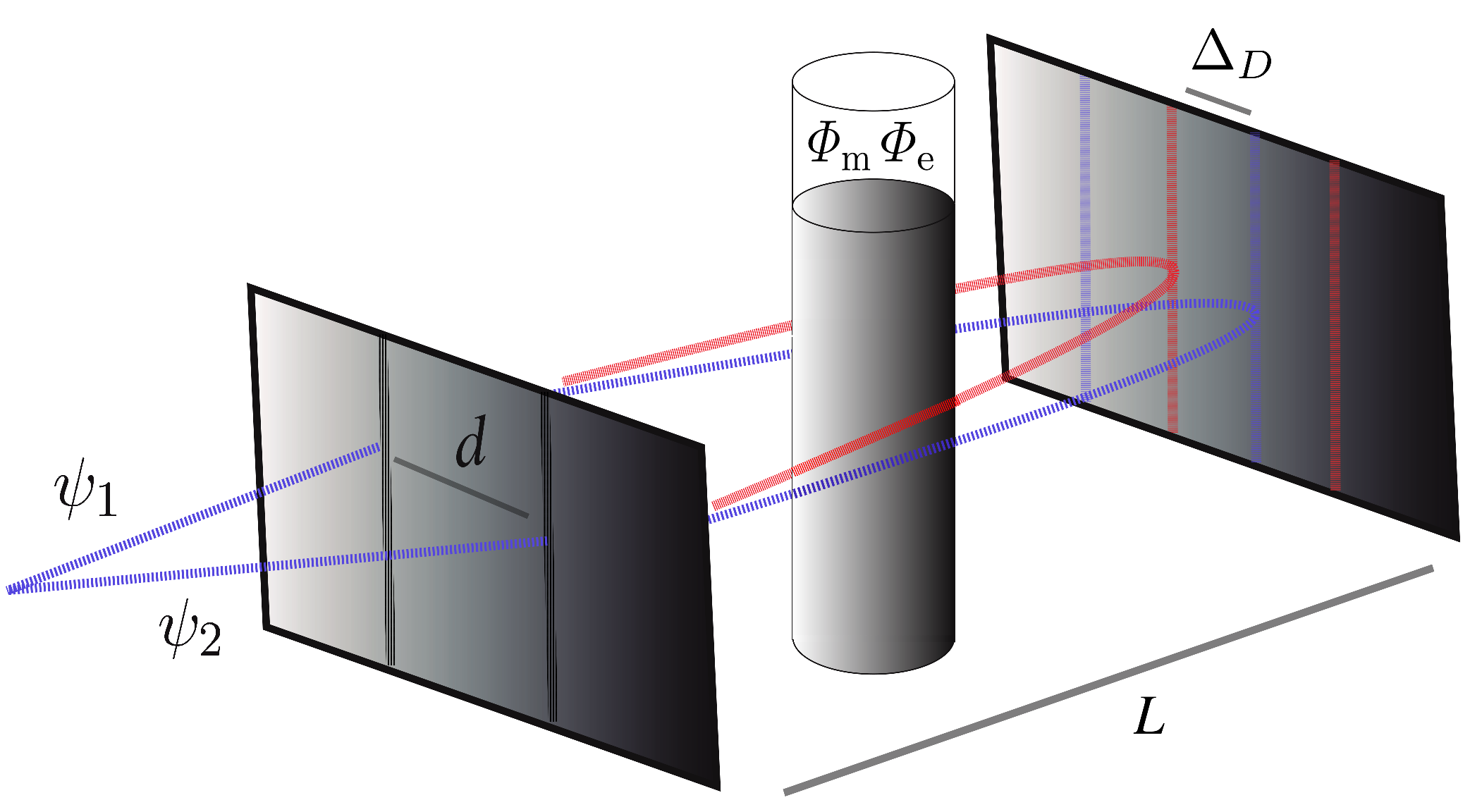}
  \caption{{\small Quantum interference effect of dyons. Dashed lines indicate dyon paths without the dual solenoid and continuous lines indicate dyon paths in the presence of the dual solenoid. The interference pattern is shifted by the amount $\Delta_{\rm D}$.}}\label{Fig4}
\end{figure}

\subsection*{\normalsize 7.3 Scattering amplitude}
Consider a particle with charge $q$ whose wave function is being scattered in the $x-y$ plane outside an infinitely-long and impenetrable cylindrical shell of radius $\cal{R}$ and centred along the $z$-axis. Inside the impenetrable cylindrical shell lies an infinitely-long solenoid of radius $R$ that confines its magnetic field. Afanasiev \cite{18} has calculated the corresponding scattering amplitude owed by the confined magnetic flux, which can be written in terms of the electromagnetic angular momentum as
\begin{equation}
f_{\rm AB}(\phi)= \sum_{\mu=-\infty}^{\infty}\frac{{\rm e}^{i \mu \phi}}{\sqrt{2\pi i k}}\Bigg[ \frac{H^{(2)}_{|\mu|}(k{\cal R})}{H^{(1)}_{|\mu|}(k{\cal R})} - {\rm e}^{i\pi(|\mu|-|\mu- L_q/\hbar|)} \frac{H^{(2)}_{|\mu-L_q/\hbar|}(k{\cal R})}{H^{(1)}_{|\mu -L_q/\hbar|}(k{\cal R})}  \Bigg],
\end{equation}
where $H^{(1)}_{|\mu|}$ is the Hankel functions of the first kind, $H^{(2)}_{|\mu|}$ is the Hankel function of the second kind, $\mu$ is an integer, and $k= \sqrt{2 M E/\hbar^2}$ is the magnitude of the wave vector with $E$ being the energy associated to the corresponding time-independent Schr\"odinger equation. Using Eq.~(48) in Eq.~(53) and inserting $L_{qg}= (q\Phi_{m}-g\Phi_{e})/(2\pi c)$, we obtain the scattering amplitude
\begin{align}
\nonumber f_{\rm D}(\phi)&= \sum_{\mu=-\infty}^{\infty}\frac{{\rm e}^{i \mu \phi}}{\sqrt{2\pi i k}}\\
&\times \Bigg[ \frac{H^{(2)}_{|\mu|}(k{\cal R})}{H^{(1)}_{|\mu|}(k{\cal R})} - {\rm e}^{i\pi(|\mu|-|\mu- (q\Phi_m-g\Phi_e)/(2\pi \hbar c)|)} \frac{H^{(2)}_{|\mu-(q\Phi_m-g\Phi_e)/(2\pi \hbar c)|}(k{\cal R})}{H^{(1)}_{|\mu -(q\Phi_m-g\Phi_e)/(2\pi \hbar c)|}(k{\cal R})} \Bigg],
\end{align}
which correspond to a dyon scattered in $x-y$ plane outside a dual solenoid (see Fig.~5), where $k=\sqrt{2 m  E /\hbar^2}$ is the magnitude of the wave vector of the dyon with $E$ denoting the energy associated to the corresponding time-independent Schr\"odinger equation. The corresponding differential scattering cross section  $d \sigma/d \Omega$ is related to scattering amplitude by $d \sigma/d \Omega= |f_{\rm D}|^2,$ where $\sigma$ is the total cross section and $\Omega$ the associated solid angle by which the dyon's wave function scatters into. Using Eq.~(54) this differential scattering cross section is given by
\begin{equation}
\frac{d \sigma}{d \Omega}= \frac{1}{2\pi k}\Bigg|\sum_{\mu=-\infty}^{\infty} \frac{H^{(2)}_{|\mu|}(k{\cal R})}{H^{(1)}_{|\mu|}(k{\cal R})} - {\rm e}^{i\pi(|\mu|-|\mu- (q\Phi_m-g\Phi_e)/(2\pi \hbar c)|)} \frac{H^{(2)}_{|\mu-(q\Phi_m-g\Phi_e)/(2\pi \hbar c)|}(k{\cal R})}{H^{(1)}_{|\mu -(q\Phi_m-g\Phi_e)/(2\pi \hbar c)|}(k{\cal R})}  \Bigg|^2.
\end{equation}
Equation (55) is admittedly cumbersome. However, we will see in the next subsection that when we assume a dual solenoid of vanishing radius $R\to 0$ the result in Eq.~(55) considerably simplifies.

\begin{figure}
  \centering
  \includegraphics[width= 200pt]{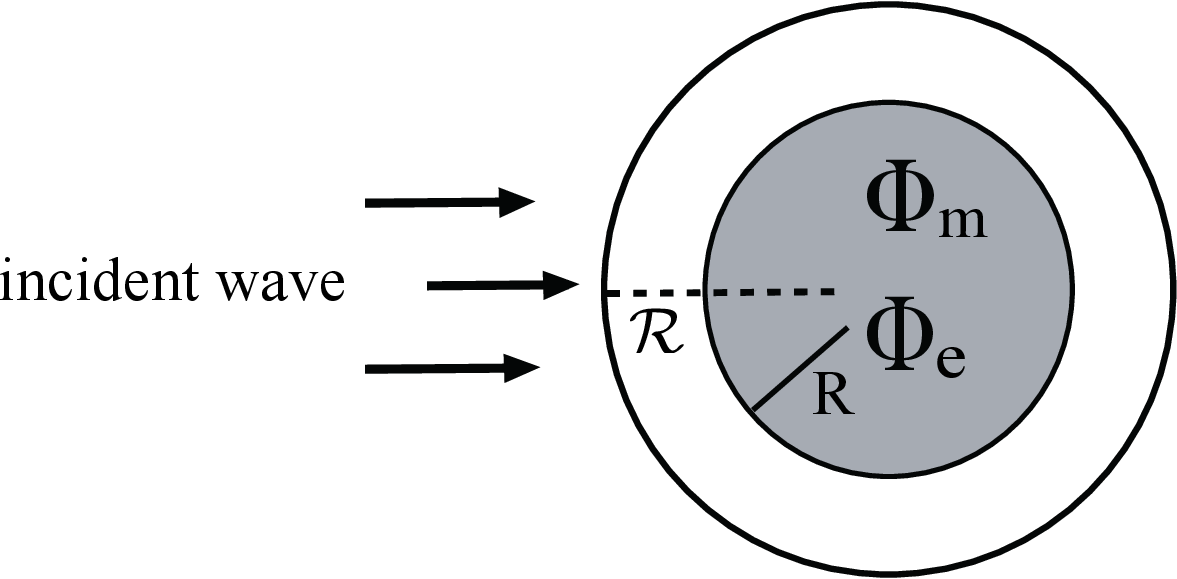}
  \caption{\small The wave function of a dyon propagates toward an infinitely-long and impenetrable cylindrical shell of radius ${\cal R}.$ Inside the cylindrical shell and infinitely-long dual solenoid of radius $R$ that confines its electric and magnetic fluxes is placed.}\label{Fig5}
\end{figure}

\subsection*{\normalsize 7.4 When the radius of the dual solenoid tends to zero: the dual flux line}
\noindent The results of this section assume that the dual solenoid has a finite radius $R$. The question then arises: are these results valid when the radius tends to zero?, i.e.,  for the case of a dual flux line. The answer is in the affirmative. Furthermore, the assumption of a dual flux line considerably simplifies the cumbersome expression in Eq.~(55) which holds for a dual solenoid of finite radius.

As $R\to 0$ the fields of the dual solenoid become confined along a line of singularity localised on the $z-$axis such that $\v B = \Phi_m \delta(\rho) \hat{\v z}/(2\pi \rho)$ and $\v E = \Phi_e \delta(\rho) \hat{\v z}/(2\pi \rho)$, where now the fluxes should be defined as $\Phi_m=4 \pi \lambda_m$ and $\Phi_e=4\pi\lambda_e,$ with $\lambda_m$ being the magnetic dipole moment density per unit length and $\lambda_e$ the electric dipole moment density per unit length. However, outside the dual flux line the corresponding vector potentials $\v A_{\rm out}$ and $\v C_{\rm out}$ have the same form as those in Eq.~(4), modulo the appropriate substitution of the corresponding constant fluxes. Accordingly, for the case of a dual flux line, the Hamiltonian operator in Eq.~(46) remains unchanged and therefore the derived quantities from this operator, such as the phase $\delta_{\rm D},$ the energy levels in Eq.~(50), the interference shift in Eq.~(52), and the scattering amplitude in Eq.~(54), all remain unchanged. In particular, for the case of the scattering amplitude, the assumption of a dual flux line considerably simplifies the cumbersome expression in Eq.~(54). In Appendix B we show that the scattering amplitude due to the wave function of a dyon scattered in the $x-y$ plane outside the dual flux line reads
\begin{equation}
f_{\rm D}(\phi)=-\frac{2 i\,{\rm e}^{-iN \phi}}{\sqrt{2 \pi i k}}\bigg[\frac{\sin[\pi (q\Phi_m-g\Phi_e)/(2\pi \hbar c)]}{{\rm e}^{i \phi}-1} \bigg],
\end{equation}
where $N$ is an integer and we have assumed $\phi \neq 0$ and $\phi \neq 2\pi$ because there is a discontinuity on these points. Using the relation $|f_{\rm D}|^2=d\sigma/d\Omega$ we obtain the differential scattering cross section
\begin{equation}
\frac{d \sigma}{d \Omega}= \frac{\sin^2[(q \Phi_{m}- g \Phi_{e})/(2 \hbar c)]}{2 \pi k \sin^2(\phi/2)}.
\end{equation}
When $g=0,$ $\Phi_{e}=0$ and $q=e,$ with $e$ being the electron's charge, and when we make the substitution $k\to k_e,$ where $k_e$ is the magnitude of the wave vector of the electron's wave function, then Eq.~(57) reduces to the reported scattering cross section for a non-relativistic electron propagating outside a magnetic flux line \cite{19,20,21,22,23}.

\section*{\large 8 On the spin of the dyon}
\noindent The idea that the dyon has spin has been discussed very little in the literature. Some authors postulate a Dirac equation for a dyon \cite{24,25,26} and thus assume that a dyon is a spin 1/2 fermion. Other authors postulate a Pauli equation for a nonrelativistic dyon \cite{27} and thus also assume that the dyon is a spin 1/2 fermion. A phenomenological model \cite{28} assumes that the spin of a dyon can correspond to that of a fermion or a boson. If we assume that a dyon is a pin 1/2 fermion, then a consistent way to introduce the spin into the Hamiltonian operator given in Eq.~(46) is as follows
\begin{equation}
\widehat{H}_{\rm D}= \frac{1}{2m}\bigg[ \bfsigma \cdot \bigg(-i \hbar \nabla - \frac{1}{c}\big(q \v A_{\rm out} + g \v C_{\rm out}\big)  \bigg) \bigg]^2 + V,
\end{equation}
where $\bfsigma=(\sigma_1, \sigma_1, \sigma_3)$ is the Pauli operator with $\sigma_i$ denoting the Pauli matrices. The corresponding Pauli equation in Dirac's bra-ket notation reads $i \hbar (\partial \ket{\Psi} / \partial t)=\widehat{H}_{\rm D}\ket{\Psi},$ where now $\ket{\Psi}$ is a two-component state function corresponding to a ``spin up" state $\Psi^{+}$ and a ``spin down" state $\Psi^{-}$. After some algebra, the Hamiltonian operator in Eq.~(58) reduces to the more explicit form
\begin{equation}
\widehat{H}_{\rm D}= \bigg[\frac{1}{2m}\bigg(-i \hbar \nabla - \frac{1}{c}\big(q \v A_{\rm out} + g \v C_{\rm out}\big)  \bigg)^2 +V\bigg]+\frac{g}{2m c} \bfsigma \cdot \v E_{\rm out}- \frac{q}{2m c}\bfsigma\cdot \v B_{\rm out}.
\end{equation}
which can be expressed as
\begin{equation}
\widehat{H}_{\rm D}= \bigg[\frac{1}{2m}\bigg(-i \hbar \nabla - \frac{1}{c}\big(q \v A_{\rm out} + g \v C_{\rm out}\big)  \bigg)^2 +V\bigg]
+\frac{2}{\gamma}\big(\,\widehat{\v d} \cdot \v E_{\rm out} - \widehat{\bfmu} \cdot \v B_{\rm out}\big),
\end{equation}
where the electric moment operator $\widehat{\v d}$ and the magnetic moment operator $\widehat{\bfmu}$ are given by
\begin{equation}
\widehat{\bfmu}= \frac{g_{\rm D}\mu}{2}\bfsigma, \quad \widehat{\v d}= \frac{g_{\rm D} d}{2}\bfsigma,
\end{equation}
with $\mu= q/(2m c)$ and $d=g/(2m c)$ being the magnetic moment and electric moments, and $\gamma$ the corresponding g-factor. Interestingly, Eq.~(60) indicates that a dyon has both an electric moment and magnetic moment, i.e., the dyon has a dual moment and in this sense, we can say that a dyon is also a dual dipole. To see how Eq.~(59) becomes Eq.~(60), we use the relation $\widehat{\v S}=(\hbar/2)\bfsigma,$ where $\widehat{\v S}$ is the spin operator,  to obtain the operators given in Eq.~(61). Using these operators we obtain
$g \bfsigma/(2mc)=2\widehat{\v d}/\gamma$ and  $q \bfsigma/(2mc)=2\widehat{\bfmu}/\gamma$ and therefore the last terms in Eqs.~(59) and (60) are the same. The right-hand side of Eq.~(60) manifests a duality between the dipole moments of the dyon and a duality between the external electric and magnetic fields. This means that the last term in Eq.~(60) is invariant under the duality transformations: $\{ \widehat{\bfmu}\to \widehat{\v d}, \,\widehat{\v d}\to -\widehat{\bfmu}\}$ and $\{\v E_{\rm out}\to \v B_{\rm out},\,\v B_{\rm out}\to -\v E_{\rm out}\}$. We note that Heras \cite{29} has discussed the duality of dipoles.

But the more important point for our purposes regarding either Eq.~(59) or Eq.~(60) is that the fields outside the dual solenoid satisfy $\v B_{\rm out}=\nabla \times \v A_{\rm out}=0$ and $\v E_{\rm out}=-\nabla \times \v C_{\rm out}=0$, and therefore the last two terms in Eq.~(59) or in Eq.~(60) identically vanish. Accordingly, if we assume the validity of Eq.~(59) or (60), then the spin of the dyon does not affect the phase $\delta_{\rm D}$. Therefore, the results obtained in Sec.~7, which are connected with this phase, remain unaffected by the spin of the dyon.

The idea that a spin 1/2 monopole should have an associated electric moment was suggested by Amaldi \cite{30}. This idea was extended by Schwinger \cite{8}, who pointed out that a spin 1/2 dyon should have both magnetic and electric moments, i.e., that a dyon should be a dual dipole as well. According to Schwinger, the electric moment of the dyon should be proportional to the product of the magnetic charge of the dyon and its spin vector, an idea supported by the second equality in Eq.~(61). More recently, Kobayashi \cite{31} showed in the context of supersymmetric Yang-Mills theories that a spin 1/2 monopole should have an electric moment. This idea was later extended by Kobayashi et al. \cite{32}, who suggested that a spin 1/2 dyon should have electric and magnetic moments. Interestingly, these authors found a similar relation to that noted by Schwinger in which the electric moment of the dyon should be proportional to the product of the magnetic charge of the dyon and its spin vector.

Finally, using Eq.~(60) and assuming the Schwinger-Zwanziger quantisation condition \cite{6,7,8}, we can argue that the electric moment of a spin 1/2 dyon is greater than its magnetic moment. Consider a dyon having the elementary electric and magnetic charges $q=e$ and $g_0=e/(2 \alpha),$ which follow from a solution of the Schwinger-Zwanziger quantisation condition,\footnote{The Schwinger-Zwanziger quantisation condition states that a pair of dyons having the charges ($q_1,g_1$) and ($q_2,g_2$) satisfy the duality-invariant relation $q_1g_2-q_2g_1= N \hbar c/2$ where $N$ is an integer. A solution of this quantisation condition is given by $q_1=(n_1)_e e, q_2=(n_2)_e e, g_1=(n_1)_g g_0,$ and $g_2=(n_2)_g g_0$ where $(n_1)_e, (n_2)_e, (n_1)_g$, and $(n_2)_g$ are integers, and $e$ and $g_0$ are the quanta of electric and magnetic charge connected by $g_0=e/(2\alpha).$ If the first dyon has the elementary charges then $q_1=e$ and $g_1=g_0=e/(2\alpha)$ which is the case we have considered. We note that Witten \cite{33} found a more general solution to the Schwinger-Zwanziger quantisation condition given by $q_1=(n_1)_e e + (n_1)_g e\theta/2\pi, q_2=(n_2)_e e +(n_2)_g e\theta/2\pi, g_1=(n_1)_g g_0,$ and $g_2=(n_2)_g g_0$, where $\theta$ is the vacuum angle, indicating that this solution assumes CP violation. Here we have not considered Witten's solution because we assume CP is conserved in our model.} where $\alpha=e^2/(\hbar c)$ is the fine structure constant. The electric and magnetic moments of this dyon are given by $\mu= e/(2mc)$ and $d = e/(4mc \alpha),$ or equivalently, $d =\mu/(2\alpha)\approx 68.5 \mu,$ indicating that the electric moment of the dyon is greater than its magnetic moment by a factor $1/(2\alpha)$. Since the charges a dyon are quantised $q=n_{e}e$ and $g=n_{g}g_0,$ with $n_e$ and $n_g$ integers, it follows that the relation $d=\mu/(2\alpha)$ holds in general. This result and Eq.~(61) connect the electric and magnetic moment operators
\begin{equation}
\widehat{\v d}= \bigg(\frac{1}{2\alpha}\bigg)\widehat{\bfmu},
\end{equation}
which states that, under the considerations made, the electric moment of a spin 1/2 dyon is greater by a factor of $(1/2 \alpha)$ compared with its corresponding magnetic moment. We emphasize that the arguments that lead to Eq.~(62) are heuristic and based on the validity of Eq.~(60) and the Schwinger-Zwanziger quantisation condition.

\section*{\large 9 Discussion: looking for dyons through the duality-invariant quantum phase}
\noindent
The story of magnetic monopoles has a beautiful part as well as a tortuous part. The beautiful part is that magnetic monopoles 
explain the observed quantisation of the electric charge through the Dirac quantisation condition \cite{34} (for a pedagogical review of this condition, see Heras \cite{35}) and are a prediction of grand unified theories \cite{36,37}. Moreover, monopoles deal with one of the most profound ideas in physics: duality. More in general, duality is a beautiful symmetry that unifies seemingly unrelated theories \cite{38,39,40}. The paradigmatic example of this symmetry is the electromagnetic duality that unifies the electrodynamics of electric and magnetic charges. The tortuous part of the story is that after more than 90 years since Dirac proposed them, all experimental efforts to detect these particles have been unsuccessful (for recent reviews, see Refs.~\cite{41,42}). In this regard, monopoles are still a sleeping beauty. However, if monopoles do not exist as isolated particles, then a generalisation of them, the so-called dyons could exist. We must say that dyons are interesting for similar reasons monopoles are: dyons explain the quantisation of the electric charge  through the Schwinger-Zwanziger quantisation condition \cite{6,7,8}, appear as predictions of grand unified theories \cite{43}, and obey the electromagnetic duality of electric and magnetic charges. However, while the experimental search for monopoles has been extensive, the experimental search for dyons has just begun \cite{28}. An exciting idea that we have developed here has consisted in connecting dyons with topological quantum phases. The possible detection of the phase $\delta_{\rm D}$ could provide indirect evidence of dyons.

\section*{\large 10 Conclusion}
Here, we have shown that quantum mechanics and electromagnetic duality predict the existence of the duality invariant quantum phase $\delta_{\rm D}=n(q\Phi_m-g\Phi_e)/(\hbar c)$. This phase is accumulated by the wave function of a dyon with charges $q$ and $g$, which encircles an infinitely-long solenoid enclosing uniform electric and magnetic fluxes $\Phi_{e}$ and $\Phi_{m}$. We have pointed out that the phase $\delta_{\rm D}$ is topological because it depends on the winding number $n$ and is nonlocal because the enclosed electric and magnetic fields act on the dyon in a region where these fields are zero. We have shown that the duality symmetry of the phase $\delta_{\rm D}$ unifies: (i) the AB phase: $\delta_{\rm AB}=nq\Phi_{m}/(\hbar c)$, which is accumulated by the wave function of an electric charge $q$ encircling an infinitely-long magnetic solenoid enclosing a uniform magnetic flux $\Phi_{m}$ and (ii) the DAB phase: $\delta_{\rm DAB}=-ng\Phi_{e}/(\hbar c)$, which is accumulated by the wave function of a magnetic charge $g$ encircling an infinitely-long electric solenoid enclosing a uniform electric flux $\Phi_{e}$. We have noted that the phase $\delta_{\rm AB}$ admits two interpretations in the light of the duality symmetry of the phase  $\delta_{\rm D}.$ We have obtained the energy levels, the two-slit interference shift, and the scattering amplitude associated with the phase $\delta_{\rm D}.$ We have showed that these results also hold for the case of a dual flux line. We have briefly commented on the role of the spin of the dyon and showed that if the dyon is a fermion of spin 1/2 then its spin does not contribute to the phase $\delta_{\rm D}$ and, therefore, to any of the results associated with this phase and discussed in  Sec.~7. In our discussion of the dyon spin, we have suggested that a spin 1/2 dyon should have both electric and magnetic moments and argued why the former should be greater than the latter.

Finally, we would like to comment on two potential applications of the quantum mechanical model proposed here in which a dyon moves around a dual solenoid. The first deals with the demonstration that the phase $\delta_{\rm D}$ is an example of the Berry/geometric phase \cite{44}, which would connect duality with geometric phases. The second deals with anyons \cite{45,46} or composites formed by an electric charge encircling a magnetic flux tube that obey fractional statistics. The dual model presented here could be used to extend the concept of anyons to composites formed by a dyon encircling a dual flux tube, which could be called duality-invariant anyons, or ``d-anyons" for short. Expectably, such d-anyons would also have fractional statistics, but their duality symmetry could bring new theoretical ideas.

\vskip 9pt
\noindent \textbf{Acknowledgements.} I thank my dad Jos\'e A. Heras for the entretaining discussions we had about the electromagnetic duality symmetry. I also thank the referee for having suggested the idea of considering the spin of the dyon.

\renewcommand\theequation{A\arabic{equation}}
\setcounter {equation}{0}
\section*{\large Appendix A: derivation of Eq.~(8)}
Consider the nonlocal Lagrangian in Eq.~(7)
\begin{equation}
L(\v x;\v x_0, \dot{\v x})=\frac{m \dot{\v x}^2}{2} + \frac{\dot{\v x}}{c}\cdot[q\v A(\v x) + g\v C(\v x)] + \int^{\v x}_{\v x_0}[g \v B(\v x) + q \v E(\v x)]\cdot d \v x',
\end{equation}
The corresponding Euler-Lagrange equations are given by
\begin{equation}
\frac{d}{dt}\bigg(\frac{\partial L}{\partial \dot{\v x}} \bigg)- \frac{\partial L}{\partial \v x}=0.
\end{equation}
Using Eq.~(A1) it follows
\begin{equation}
\frac{\partial L}{\partial \dot{\v x}} = m \dot{\v x} + \frac{1}{c}(q \v A + g \v C),
\end{equation}
so that the first term in the Euler-Lagrange equations reads
\begin{equation}
\frac{d}{dt}\bigg(\frac{\partial L}{\partial \dot{\v x}} \bigg)= m \ddot{\v x} + \frac{1}{c}\frac{d}{dt}(q \v A + g \v C).
\end{equation}
Using the result $(d\v F/dt)=(\dot{\v x}\cdot \nabla)\v F$ which holds for a time-independent vector function $\v F = \v F(\v x)$ it follows $(d/dt)(q \v A_{\rm out} + g \v C_{\rm out})=(\dot{\v x} \cdot \nabla)(q \v A + g \v C)$ which is used in Eq.~(A4) to obtain
\begin{equation}
\frac{d}{dt}\bigg(\frac{\partial L}{\partial \dot{\v x}} \bigg)= m \ddot{\v x} + \frac{1}{c}(\dot{\v x}\cdot \nabla)(q \v A + g \v C).
\end{equation}
On the other hand, from the identity $\dot{\v x}\times (\nabla \times \v F)=\nabla(\dot{\v x}\cdot \v F)-(\v F\cdot \nabla)\dot{\v x},$ it follows $(\dot{\v x}\cdot \nabla)(q \v A + g \v C)=-\dot{\v x}\times \nabla \times (q \v A + g \v C)+\nabla[\dot{\v x} \cdot (q\v A + g \v C)],$ which is used in Eq.~(A5) to obtain
\begin{equation}
\frac{d}{dt}\bigg(\frac{\partial L}{\partial \dot{\v x}} \bigg)= m \ddot{\v x} -\frac{\dot{\v x}}{c}\times \nabla \times(q \v A + g \v C) + \frac{1}{c}\nabla[\dot{\v x} \cdot (q \v A + g \v C)].
\end{equation}
The second term in the Euler-Lagrange equations read
\begin{equation}
\frac{\partial L}{\partial \v x} = \frac{1}{c}\nabla[\dot{\v x}\cdot (q \v A + g\v C)]+\nabla \bigg[ \int^{\v x}_{\v x_0}[g \v B(\v x) + q \v E(\v x)]\cdot d \v x'\bigg]
\end{equation}
Let us evaluate the second term. The line integral gives
\begin{equation}
\int^{\v x}_{\v x_0}[g \v B(\v x) + q \v E(\v x)]\cdot d \v x'= [g B + q E](z-z_0),
\end{equation}
where $B= \Phi_m\Theta(R-\rho)/(\pi R^2)$ and $E=\Phi_e \Theta(R-\rho)/(\pi R^2)$. Therefore
\begin{eqnarray}
\nonumber \nabla\bigg[\int^{\v x}_{\v x_0}[g \v B(\v x) + q \v E(\v x)]\cdot d \v x' \bigg] &=& \nabla [(g B + q E)(z-z_0)]\\
\nonumber &=& \frac{(g\Phi_m + q \Phi_e)}{\pi R^2}\nabla[\Theta(R-\rho)(z-z_0)]\\
\nonumber &=& \frac{(g\Phi_m + q \Phi_e)}{\pi R^2}[ (z-z_0)\delta(\rho- R)\,\hat{\!\bfrho} + \Theta(R-\rho)\hat{\v z}]\\
\nonumber &=& g \bigg [\frac{\Phi_m \Theta(R-\rho) }{\pi R^2} \hat{\v z}\bigg] + q \bigg [\frac{\Phi_e \Theta(R-\rho) }{\pi R^2} \hat{\v z}\bigg] \\
\nonumber &-& \bigg[\frac{(g\Phi_m + q \Phi_e)}{\pi R^2} (z_0-z)\delta(\rho- R)\,\hat{\!\bfrho} \bigg]\\
&=& g \v B + q \v E - \v F_{\rm s},
\end{eqnarray}
where $\v E$ and $\v B$ are the fields defined by Eq.~(2) and $\v F_{\rm s}$ is the singular term defined in Eq.~(9). Using Eq.~(A9) in Eq.~(A7) it follows
\begin{equation}
\frac{\partial L}{\partial \v x} = \frac{1}{c}\nabla[\dot{\v x}\cdot (q \v A + g\v C)]+ g \v B +q \v E +\v F_{\rm s}.
\end{equation}
Using Eqs.~(A6) and (A10) we obtain
\begin{equation}
\frac{d}{dt}\bigg(\frac{\partial L}{\partial \dot{\v x}} \bigg)-\frac{\partial L}{\partial \v x}=m \ddot{\v x} -\frac{\dot{\v x}}{c}\times \nabla \times(q \v A + g \v C)-g \v B - q \v E - \v F_{\rm s}=0.
\end{equation}
Identifying $\nabla\times \v A = \v B$, $-\nabla \times \v C=\v E$, and $\v F= m \ddot{\v x}$ it follows
\begin{equation}
\frac{d}{dt}\bigg(\frac{\partial L}{\partial \dot{\v x}} \bigg)-\frac{\partial L}{\partial \v x}=\v F -\bigg[q\bigg(\v E + \frac{\dot{\v x}}{c}\times \v B \bigg) + g\bigg(\v B - \frac{\dot{\v x}}{c} \times \v E\bigg)\bigg] - \v F_{\rm s}=0.
\end{equation}
Equation (A12) implies
\begin{equation}
\v F=\bigg[q\bigg(\v E + \frac{\dot{\v x}}{c}\times \v B \bigg) + g\bigg(\v B - \frac{\dot{\v x}}{c} \times \v E\bigg)\bigg] + \v F_{\rm s},
\end{equation}
which is Eq.~(8).

\renewcommand\theequation{B\arabic{equation}}
\setcounter {equation}{0}
\section*{\large Appendix B: Scattering amplitude with a dual flux line}
Consider the scattering amplitude in Eq.~(54)
\begin{align}
\nonumber f_{\rm D}(\phi)&= \sum_{\mu=-\infty}^{\infty}\frac{{\rm e}^{i \mu \phi}}{\sqrt{2\pi i k}}\\
&\times\Bigg[ \frac{H^{(2)}_{|\mu|}(k{\cal R})}{H^{(1)}_{|\mu|}(k{\cal R})} - {\rm e}^{i\pi(|\mu|-|\mu- (q\Phi_m-g\Phi_e)/(2\pi \hbar c)|)} \frac{H^{(2)}_{|\mu-(q\Phi_m-g\Phi_e)/(2\pi \hbar c)|}(k{\cal R})}{H^{(1)}_{|\mu -(q\Phi_m-g\Phi_e)/(2\pi \hbar c)|}(k{\cal R})}  \Bigg],
\end{align}
corresponding to a dyon scattered in the $x-y$ plane outside a dual solenoid of finite radius $R$ enclosed by an impenetrable cylindrical shell of radius $\cal{R}.$ We want to find the form of Eq.~(B1) as $R\to 0,$ i.e., when the dual solenoid reduces to a dual flux line. We will now follow arguments due to Afanasiev \cite{18} and Hagen \cite{47}.

In the limit $R\to 0$ we have  $k {\cal R}<<1$ and the ratio of the Hankel functions of the second and first kind in Eq.~(B1) can be approximated to \cite{18}
\begin{equation}
\frac{H^{(2)}_{|\mu|}(k{\cal R})}{H^{(1)}_{|\mu|}(k{\cal R})}\approx- 1, \quad \frac{H^{(2)}_{|\mu-(q\Phi_m-g\Phi_e)/(2\pi \hbar c)|}(k{\cal R})}{H^{(1)}_{|\mu -(q\Phi_m-g\Phi_e)/(2\pi \hbar c)|}(k{\cal R})}\approx - 1.
\end{equation}
Inserting Eq.~(B2) in Eq.~(B1) it reduces to
\begin{equation}
f_{\rm D}(\phi)= \sum_{\mu=-\infty}^{\infty}\frac{{\rm e}^{i \mu \phi}}{\sqrt{2 \pi i k}}\bigg[ {\rm e}^{ 2 i \delta_\text{s}} -1\bigg],
\end{equation}
where $\delta_{\rm s}$ is the scattering phase given by
\begin{equation}
\delta_{\rm s}= \frac{\pi}{2}|\mu| -\frac{\pi}{2}\bigg|\mu-\frac{(q\Phi_m-g\Phi_e)}{2\pi \hbar c}\bigg|.
\end{equation}
To compute the summation in Eq.~(B3) we let $[(q\Phi_m-g\Phi_e)/(2\pi \hbar c)]= N + \beta$ where $N$ is the largest integer less than $ [(q\Phi_m-g\Phi_e)/(2\pi \hbar c)]$ and the quantity $\beta$ is any non-integer in the range $0\leq\beta<1.$ Using this result we can write the scattering phase in Eq.~(B4) as $\delta_{\rm s}=(\pi/2)|\mu|-(\pi/2)|\mu - N - \beta|$. Now, if $\mu\geq N$ then $|\mu|- |\mu- N-\beta|= [(q\Phi_m-g\Phi_e)/(2\pi \hbar c)]$ and if $\mu<-N$ then $|\mu|- |\mu- N-\beta|= -[(q\Phi_m-g\Phi_e)/(2\pi \hbar c)].$ Therefore, we can write the scattering phase in Eq.~(B4) as
\begin{equation}
\delta_{\rm s}=
\begin{cases}
-\dfrac{\pi}{2} \dfrac{(q\Phi_m-g\Phi_e)}{2\pi \hbar c}, &  \mu \geq N\\[0.5cm]
+\dfrac{\pi}{2} \dfrac{(q\Phi_m-g\Phi_e)}{2\pi \hbar c}, & \mu<-N
\end{cases}
\end{equation}
Using Eq.~(B5) in Eq.~(B3) and after an appropriate handling of the summation limits we obtain
\begin{align}
\nonumber f_{\rm D}(\phi)&=\frac{1}{\sqrt{2 \pi i k}}\\
&\times\Bigg[\sum_{\mu=-N}^{\infty} {\rm e}^{i \mu \phi}(  {\rm e}^{i\pi(q\Phi_m-g\Phi_e)/(2\pi \hbar c) } -1) + \sum_{\mu=-\infty}^{- N -1}  {\rm e}^{i \mu\phi}({\rm e}^{-i\pi (q\Phi_m-g\Phi_e)/(2\pi \hbar c)} -1) \Bigg].
\end{align}
The summations in Eq.~(B6) are now simple to evaluate and they give
\begin{eqnarray}
\sum_{\mu=-N}^{\infty} {\rm e}^{i \mu \phi}(  {\rm e}^{i\pi(q\Phi_m-g\Phi_e)/(2\pi \hbar c) } -1)=-\frac{({\rm e}^{-i N \phi})({\rm e}^{i\pi (q\Phi_m-g\Phi_e)/(2\pi \hbar c)} -1)}{({\rm e}^{i \phi }-1)},\\
\sum_{\mu=-\infty}^{- N -1}  {\rm e}^{i \mu\phi}({\rm e}^{-i\pi (q\Phi_m-g\Phi_e)/(2\pi \hbar c)} -1)=\frac{({\rm e}^{-i N \phi})({\rm e}^{-i\pi(q\Phi_m-g\Phi_e)/(2\pi \hbar c) } -1)}{({\rm e}^{i \phi }-1)},
\end{eqnarray}
where we have assumed $\phi \neq 0$ and $\phi \neq 2\pi$. Inserting the Eq.~(B7) and Eq.~(B8) in Eq.~(B6) it follows that the scattering amplitude in Eq.~(B6) reduces to
\begin{equation}
f_{\rm D}(\phi)=-\frac{2 i\,{\rm e}^{-iN \phi}}{\sqrt{2 \pi i k}}\bigg[\frac{\sin[\pi (q\Phi_m-g\Phi_e)/(2\pi \hbar c)]}{{\rm e}^{i \phi}-1} \bigg],
\end{equation}
which is Eq.~(56). Using Eq.~(B9) and $d\sigma/d\Omega=|f_{\rm D}|^2$ we obtain the corresponding differential scattering cross section in Eq.~(57).


\begin{thebibliography}{0}

\bibitem{1} Y. Aharonov and D. Bohm, \textit{Significance of Electromagnetic Potentials in the Quantum Theory}, \href{https://doi.org/10.1103/PhysRev.115.485}{Phys. Rev. \textbf{115}, 485 (1959)}.

\bibitem{2} J. P. Dowling, C. P. Williams and J. D. Franson, \textit{Maxwell Duality, Lorentz Invariance, and Topological Phase}, \href{https://doi.org//10.1103/PhysRevLett.83.2486}{Phys. Rev. Lett. \textbf{83}, 2486 (1999)}.

\bibitem{3} C. Furtado and G. Duarte, \textit{Dual Aharonov–Bohm effect}, \href{https://doi.org//10.1088/0031-8949/71/1/001}{Phys. Scr. \textbf{71}, 7 (2005)}.

\bibitem{4} Y. Aharonov and A. Casher, \textit{Topological Quantum Effects for Neutral Particles,} \href{https://doi.org/10.1103/PhysRevLett.53.319}{Phys. Rev.
Lett. \textbf{53}, 319 (1984)}.


\bibitem{5} D. Rohrlich, \textit{Duality in the Aharonov–Casher and Aharonov–Bohm effects,}
\href{https://doi.org/10.1088/1751-8113/43/35/354028}{J. Phys. A, \textbf{43}, 354028 (2010)}.


\bibitem{6}
J. Schwinger, \emph{Sources and Magnetic Charge,} \href{https://doi.org/10.1103/PhysRev.173.1536}{Phys Rev. \textbf{173}, 1536 (1968)}.

\bibitem{7}
D. Zwanziger,\emph{Quantum Field Theory of Particles with Both Electric and Magnetic Charges,} \href{https://doi.org/10.1103/PhysRev.176.1489}{Phys Rev. \textbf{176}, 1489 (1968)}.

\bibitem{8}
J. Schwinger, \emph{A Magnetic Model of Matter,} \href{https://doi.org/10.1126/science.165.3895.757}{Science \textbf{165}, 757 (1969)}.


\bibitem{9}
J. A. Heras and R. Heras, \textit{Can classical electrodynamics predict nonlocal effects?} \href{https://doi.org/10.1140/epjp/s13360-021-01835-9}{Eur. Phys. J. Plus \textbf{136}, 847 (2021)}.

\bibitem{10}
J. A. Heras and R. Heras, \textit{Topology, nonlocality and duality in classical electrodynamics}, \href{https://doi.org/10.1140/epjp/s13360-022-02364-9}{Eur. Phys. J. Plus \textbf{137}, 157 (2022)}.

\bibitem{11}
J. Maeda and K. Shizuya, \textit{The Aharonov-Bohm and Aharonov-Casher effects and electromagnetic angular momentum}, \href{https://doi.org/10.1007/BF01474622}{Z. Phys. C \textbf{60}, 265 (1993)}.

\bibitem{12}
M. Wakamatsu et al., \textit{The role of electron orbital angular momentum in the Aharonov-Bohm effect revisited},  \href{https://doi.org/10.1016/j.aop.2018.08.010}{Ann. Phys. \textbf{38}, 259 (2018)}.

\bibitem{13}
F. Rohrlich, \textit{Classical Theory of Magnetic Monopoles},
\href{https://doi.org/10.1103/PhysRev.150.1104}{Phys. Rev. \textbf{150}, 1104 (1966)}.

\bibitem{14}
D. Rosenbaum, \textit{Proof of the Impossibility of a Classical Action Principle for Magnetic Monopoles and Charges without Subsidiary Conditions}, \href{https://doi.org/10.1103/PhysRev.147.891}{Phys. Rev. \textbf{147}, 891 (1966)}.

\bibitem{15}
R. Heras, \textit{The Aharonov–Bohm effect in a closed flux line}, \href{https://doi.org/10.1140/epjp/s13360-022-02832-2}{Eur. Phys. J. Plus \textbf{137}, 641 (2022)}.

\bibitem{16}
H. Kleinert,  \textit{Multivalued Fields in Condensed Matter, Electromagnetism, and Gravitation} (World Scientific, Singapore, 2008).

\bibitem{17}
D. H. Kobe, \emph{Aharonov-Bohm effect revisited,} \href{https://doi.org/10.1016/0003-4916(79)90344-0}{ Ann. Phys. \textbf{123,} 381 (1979)}.

\bibitem{18}
G. N. Afanasiev, \textit{Topological Effects in Quantum Mechanics} (Springer, Netherlands, 1999).


\bibitem{19}
W. C. Henneberger, \emph{Some aspects of the Aharonov-Bohm effect,} \href{https://doi.org/10.1103/PhysRevA.22.1383}{Phys. Rev. A \textbf{22}, 1383 (1980)}.

\bibitem{20}
S. N. M. Ruijsenaars, \emph{The Aharonov-Bohm effect and scattering theory,} \href{https://doi.org/10.1016/0003-4916(83)90051-9}{Ann. Phys. \textbf{146}, 1 (1983)}.

\bibitem{21}
S. Sakoda and M. Omote, \textit{Aharonov–Bohm scattering: The role of the incident wave,}  \href{https://doi.org/10.1063/1.531888}{J. Math. Phys. \textbf{38}, 716 (1997)}.

\bibitem{22}
S. Sakoda and M. Omote, \textit{Difference in the Aharonov–Bohm Effect on Scattering States and Bound States,}  \href{https://doi.org/10.1016/S1076-5670(08)70209-X}{Advances in Imaging and Electron Physics, \textbf{110}, 101 (1999)}.

\bibitem{23}
U. Camara da Silva, \textit{Renormalization group flow of the Aharonov–Bohm scattering amplitude,}  \href{https://doi.org/10.1016/j.aop.2018.09.001}{Ann. Phys. \textbf{398}, 38 (2018)}.


\bibitem{24}
D. Zwanziger, \emph{Local-Lagrangian Quantum Field Theory of Electric and Magnetic Charges,} \href{https://doi.org/10.1103/PhysRevD.3.880}{Phys. Rev. D \textbf{3}, 880 (1971)}.

\bibitem{25}
S. G. Kovalevich, et al., \textit{Effective Lagrangian of QED with a magnetic charge and dyon mass bounds,} \textit{Effective Lagrangian of QED with a magnetic charge and dyon mass bounds,}  \href{https://doi.org/10.1103/PhysRevD.55.5807}{Phys. Rev. D \textbf{55}, 5807 (1997)}.

\bibitem{26}
Y. M. Shnir, \emph{Magnetic monopoles,} (Springer, Berlin, 2005).


\bibitem{27}
P. C. Pant, V. P. Pandey, and B. S. Rajput, \textit{Pauli equation for spin-1/2 dyonium in non-Abelian gauge theory}, \href{https://doi.org/10.1007/BF03185565}{Il Nuovo Cimento A (1971-1996) \textbf{110}, 1421 (1997)}.

\bibitem{28}
B. Acharya et al. (MoEDAL Collaboration), \textit{First Search for Dyons with the Full MoEDAL Trapping Detector in 13 TeV $\textbf{pp}$ Collisions,} \href{https://doi.org/10.1103/PhysRevLett.126.071801}{Phys. Rev. Lett. \textbf{126}, 071801, (2021)}.


\bibitem{29}
J. A. Heras, \textit{Unified model based on U(1) duality symmetry of polarization and magnetization},  \href{https://doi.org/10.1103/PhysRevE.58.R6951}{Phys. Rev. E \textbf{58}, R6951(R) (1998)}.

\bibitem{30}
E. Amaldi, \textit{On the Dirac Magnetic Poles}, in Old and New Problems in. Elementary Particles, ed., G. Puppi, Academic Press, New York (1968).


\bibitem{31}
M. Kobayashi, \textit{Electric Dipole Moment of Magnetic Monopole}, \href{https://doi.org/10.1143/PTP.117.479}{Prog. Theor. Phys. \textbf{117}, 479 (2007)}.


\bibitem{32}
M. Kobayashi, T. Kugo, and T. Tokunaga, \textit{Electric Dipole Moments of Dyon and `Electron,’} \href{https://doi.org/10.1143/PTP.118.921}{Prog. Theor. Phys. \textbf{118}, 921 (2007)}.


\bibitem{33}
E. Witten, \emph{Dyons of charge $e\theta/2\pi,$} \href{https://doi.org/10.1016/0370-2693(79)90838-4}{Phys. Lett. B \textbf{86}, 283 (1979)}.

\bibitem{34}
P.A.M. Dirac, \textit{Quantised singularities in the electromagnetic field}, \href{https://doi.org/10.1098/rspa.1931.0130}{Proc. R. Soc. Lond. A. \textbf{133}, 60 (1931)}.

\bibitem{35}
R. Heras, \emph{Dirac quantisation condition: a comprehensive review,} \href{https://doi.org/10.1080/00107514.2018.1527974}{Contemp. Phys. \textbf{59}, 331 (2018)}.

\bibitem{36}
G.'t Hooft, \textit{Magnetic monopoles in unified gauge theories},  \href{https://doi.org/10.1016/0550-3213(74)90486-6}{ Nucl. Phys. B \textbf{79}, 276 (1974)}.

\bibitem{37}
A. M. Polyakov, \textit{Particle Spectrum in the Quantum Field Theory,} JETP Letters \textbf{20,} 194 (1974).

\bibitem{38}
P. Townsend, \textit{Unity from duality}, \href{https://doi.org/10.1088/2058-7058/8/9/24}{Phys. World \textbf{8}, 41 (1995)}.

\bibitem{39}
D. Olive,  \textit{Exact electromagnetic duality,} \href{https://doi.org//10.1016/0920-5632(95)00618-4}{Nucl. Phys. B (Proc. Suppl) \textbf{46}, 88 (1996)}.

\bibitem{40}
E. Witten, \textit{Duality, Spacetime and Quantum Mechanics,} \href{https://doi.org// 10.1063/1.881616}{ Phys. Today  \textbf{50}, 28 (1997)}.


\bibitem{41}
M. Spurio, \textit{Searches for Magnetic Monopoles and Other Stable Massive Particles,
in Probing Particle Physics with Neutrino Telescopes,} (World Scientific, Singapore), ch. 11, pp. 353–400.


\bibitem{42}
N. E. Mavromatos and V. A. Mitsou, \textit{Magnetic monopoles revisited: Models and searches at colliders and in the Cosmos}, \href{https://doi.org/10.1142/S0217751X20300124}{Int. J. Mod. Phys. A \textbf{35}, 2030012 (2020)}.

\bibitem{43}
B. Julia and A. Zee, \textit{Poles with both magnetic and electric charges in non-Abelian gauge theory,}  \href{https://doi.org/10.1103/PhysRevD.11.2227}{Phys. Rev. D \textbf{11}, 2227 (1975)}.


\bibitem{44}
M. V. Berry, \emph{Quantal phase factors accompanying adiabatic changes,} \href{https://doi.org/10.1098/rspa.1984.0023}{Proc. R. Soc. Lond. A. \textbf{392}, 45 (1984)}.

\bibitem{45}
G. A. Goldin, \textit{Representations of a local current algebra in nonsimply connected space and the Aharonov–Bohm effect,}  \href{https://doi.org/10.1063/1.525110}{J. Math. Phys. \textbf{22}, 1664 (1981)}.


\bibitem{46}
F. Wilczek, \textit{Quantum Mechanics of Fractional-Spin Particles}, \href{https://doi.org/10.1103/PhysRevLett.49.957}{Phys. Rev. Lett. \textbf{49}, 957 (1982)}.

\bibitem{47}
C. R. Hagen, \textit{Aharonov-Bohm scattering amplitude},  \href{https://doi.org/10.1103/PhysRevD.41.2015}{Phys. Rev. D \textbf{41}, 2015 (1990)}.




\end{thebibliography}
\end{document}